\documentclass[prb,aps,twocolumn,amsmath,amssymb,superscriptaddress,floatfix,longbibliography]{revtex4-2}
\usepackage[breaklinks=true,colorlinks,citecolor=blue,linkcolor=blue,urlcolor=blue]{hyperref}
\usepackage{float}
\usepackage{bm}
\usepackage[normalem]{ulem}
\usepackage{epsfig,mathrsfs,color,latexsym,subfigure,marginnote,graphicx,verbatim,mathrsfs,color,array,amsmath,amsfonts,amssymb,graphicx}

\usepackage{tabularx, array}
\newcolumntype{Y}{>{\centering\arraybackslash}X}

\usepackage[utf8]{inputenc}
\usepackage{physics}

\makeatletter
\def\maketitle{
\@author@finish
\title@column\titleblock@produce
\suppressfloats[t]}
\makeatother

\begin{document}
\title{Emergent spin Hall quantization and high-order van Hove singularities in square-octagonal MA$_2$Z$_4$}

\author{Rahul Verma}
\affiliation{Department of Condensed Matter Physics and Materials Science, Tata Institute of Fundamental Research, Colaba, Mumbai 400005, India}

\author{Yash Vardhan}
\affiliation{Department of Condensed Matter Physics and Materials Science, Tata Institute of Fundamental Research, Colaba, Mumbai 400005, India}

\author{Hsin Lin}
\affiliation{Institute of Physics, Academia Sinica, Taipei 11529, Taiwan}

\author{Bahadur Singh}
\email{Contact author: bahadur.singh@tifr.res.in}
\affiliation{Department of Condensed Matter Physics and Materials Science, Tata Institute of Fundamental Research, Colaba, Mumbai 400005, India}

\begin{abstract}
Quantum spin Hall (QSH) insulators are versatile platforms for exploring exotic quantum phases, especially when combined with high-order van Hove singularities (VHSs) that enhance electron correlations. However, perfect spin Hall quantization is often hindered by spin mixing from strong spin-orbit coupling, and the emergence of such VHSs is highly sensitive to material-specific electronic structures. Here, we predict a class of seven-layered square-octagonal MA$_2$Z$_4$ (M = Mo/W, A = Si/Ge, Z = Pnictogen) isomers that host a robust, large-gap QSH phase with nearly quantized spin Hall conductivity and intrinsic high-order VHSs. Topological and symmetry analyses reveal that compounds with Z = P, As, and Sb are $\mathbb{Z}_2$ nontrivial with spin Chern number $C_s = 1$ and support $S_z$-polarized edge states, while those with Z = N are trivial insulators. The QSH phase features an $S_z$-conserving spin Hamiltonian consistent with an emergent spin $\mathrm{U}(1)$ quasi-symmetry, yielding spin Hall conductivity $\sim 2e^2/h$. Notably, MA$_2$(As, Sb)$_4$ compounds exhibit quasi-flat bands near the Fermi level in the inverted regime, with WSi$_2$Sb$_4$ additionally hosting four high-order VHSs at generic momentum points. These results position square-octagonal MA$_2$Z$_4$ materials as robust QSH insulators for realizing quantized spin Hall conductivity and correlated topological phases, including fractionalized states and possibly non-Abelian anyons.
\end{abstract}

\maketitle 
\textbf{Introduction.}
Search for nontrivial quantum phases in realistic materials remains a central pursuit in modern condensed matter physics~\cite{AM_Singh2022,RMP_Hasan2010,RMP_Bansil2016,NR_Hasan2021,NR_Xiao2021,Tokura2017}. Among these, quantum spin Hall (QSH) insulators, or two-dimensional (2D) topological insulators, remain at the forefront owing to their spin-momentum-locked helical edge modes coexisting with an insulating bulk. These systems possess a nontrivial $\mathbb{Z}_2$ invariant which, in the presence of spin-$S_z$-conserving Hamiltonians, gives rise to quantized spin Hall conductivity (SHC)~\cite{QSH_KM2005,QSH_KM2005b,HgTe_Bernevig2006,HgTe_Konig2007}. In real materials, however, spin-mixing induced by strong spin-orbit coupling (SOC) breaks spin $U(1)$ symmetry along natural crystallographic axes, causing deviations from perfect SHC quantization~\cite{QSH_KM2005,QSH_KM2005b,SCN_PRB2009,SHC_PRB2019,SHC_PRB2020}. For instance, in WTe$_2$ with $\mathbb{Z}_2 = 1$, edge conductance along the normal surface is suppressed due to a tilt of the spin axis by $(40 \pm 2)^\circ$ from the layer normal, reflecting spin-locking misaligned from the crystal axes~\cite{SC_PRX2021,SC_PRL2020,SC_NL2021}. Recent theoretical work suggests that an emergent spin $U(1)$ quasi-symmetry may enable near-quantized SHC in experimentally accessible geometries and materials, despite the absence of exact spin conservation~\cite{QS_NP2022,QS_PRL2024,QS_PRB2024,QS_SbVerma2024}. Yet, experimental realization of quantized SHC remains limited to a few systems - such as HgTe/CdTe, InAs/GaSb quantum wells, moiré WSe$_2$, and TaIrTe$_4$ - underscoring the need for improved material platforms and more reliable identification strategies for ideal QSH insulators~\cite{HgTe_Konig2007,QSH_InAs2015,QSH_T100K2018,QSH_Moire2024,QSH_moireWSe22024, DQSH_TaIrTe42024}.

Beyond single-particle topological features, incorporating strong electron correlations into QSH systems offers a promising route to realizing exotic quantum phases, including fractionalized Chern insulators and interaction-driven topological orders~\cite{Tokura2022,Cai2023,Kondo_PRL2010,SmB6_Singh2022,MWSM_BSPRL2017,CP_RPMSingh2024,CP_CPSingh2023,Hu2020,Kagome_Patra2025}. However, correlated QSH insulators are rare, as most known examples derive their topology from highly dispersive $s$ or $p$ bands, which lack van Hove singularities (VHSs) or flat dispersion near the Fermi level. A viable strategy is to identify systems where band inversion involves states with strong Coulomb interactions - such as localized $d$ or $f$ orbitals, and where the electronic structure hosts flat bands or VHSs, as seen in kagome or other bipartite lattices~\cite{SmB6_Singh2022,Kagome_Patra2025}. A notable example is TaIrTe$_4$, where a density-tuned correlated QSH phase has been reported, possibly mediated by a charge density wave associated with VHSs~\cite{DQSH_TaIrTe42024}. QSH systems that combine topological order, flat dispersion, and strong interactions may support both robust SHC and emergent correlation-driven phenomena under experimentally accessible conditions.

In this work, we identify a new family of QSH insulators featuring nearly quantized SHC and high-order VHSs in monolayers MA$_2$Z$_4$ (M = Mo/W; A = Si/Ge; Z = Pnictogen) with a square-octagonal (SO) geometry. MoSi$_2$N$_4$, recently synthesized via a bottom-up approach without bulk analogs in a 1H structure, has been shown to exhibit excellent stability and tunable electronic and spintronic properties~\cite{Hong2020,Novesolov2024}. Its symmetry-lowered 1T$^\prime$ phase hosts a QSH state with a $d-p$ band inversion but with SHC values that deviate significantly from the quantized limit of $2e^2/h$~\cite{QSH_Verma2022,Islam2023}. Using first-principles calculations, phonon spectra, and molecular dynamics, we design a seven-layered SO polymorph of MA$_2$Z$_4$ that realizes QSH phases distinct from known 1T$^\prime$ isomers. While MA$_2$N$_4$ remains topologically trivial, heavier pnictogens (Z = P, As, Sb) induce a $d-d$ band inversion at the $\Gamma$ point, accompanied by flatter quadratic band dispersions; SOC then drives the system into a topologically nontrivial QSH phase. Spin-resolved feature spectrum analysis reveals an approximate $S_z$-conserving spin Hamiltonian with emergent spin $U(1)$ quasi-symmetry, enabling a near-quantized SHC $\sim 2e^2/h$. Importantly, MA$_2$(As, Sb)$_4$ features quasi-flat dispersion, with WSi$_2$Sb$_4$ additionally hosting multiple high-order VHSs at generic momentum points near the Fermi level. The coexistence of a QSH phase with emergent spin $U(1)$ quasi-symmetry and high-order VHSs positions SO MA$_2$Z$_4$ isomers as compelling platforms for exploring the interplay between quantized SHC and correlation-driven topological phenomena.

\textbf{Methods.} Electronic structure calculations were carried out using density functional theory (DFT) within the projector augmented-wave (PAW) formalism, as implemented in the Vienna ab initio simulation package (VASP)~\cite{Hohen1964,Kresse1996,Kresse1999}. A plane-wave cutoff of 420 eV and Gaussian smearing with a width of 50 meV were applied. The generalized gradient approximation (GGA)~\cite{perdew1996generalized} was used for exchange–correlation interactions, and SOC was included self-consistently. Brillouin zone sampling employed a $\Gamma$-centered $9 \times 9 \times 1$ $k$-mesh with an energy convergence threshold of $10^{-6}$ eV. A vacuum spacing of 12~$\text{\AA}$ was added to eliminate interlayer interactions. Phonon spectra were computed using the frozen-phonon method with a $3 \times 3 \times 1$ supercell via the \textsc{Phonopy} code~\cite{phonopy}, and {\it ab initio} molecular dynamics simulations were performed using a Nosé-Hoover thermostat at 300 K with a 1 fs time step~\cite{PhysRevB.48.2081}. Tight-binding Hamiltonians were constructed using the VASP2WANNIER90 interface~\cite{mostofi2008wannier90} and employed to evaluate topological properties. Surface state spectra were obtained via the semi-infinite Green’s function approach~\cite{Wtools,Sancho1985},  and spin-resolved feature spectrum and topological analyses were performed using an in-house developed code.

\begin{figure}[t!]
\includegraphics[width=1\linewidth]{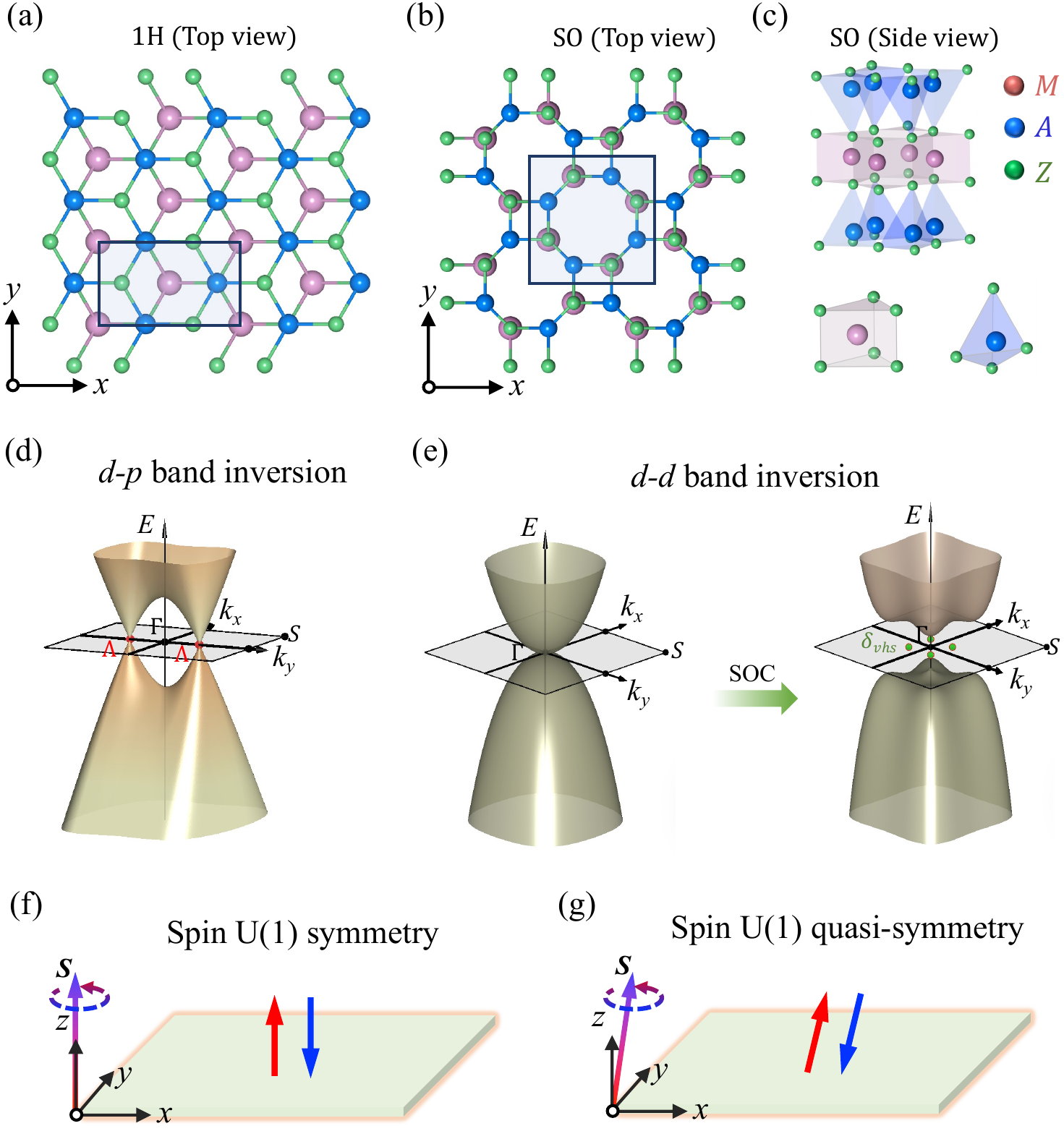}
\caption{\textbf{Atomic structure and electronic features of MA$_2$Z$_4$.} (a) Hexagonal 1H and (b) square-octagonal (SO) phases, with unit cells marked in black. (c) Side view of the SO phase showing Z–A–Z–M–Z–A–Z stacking, analogous to the 1H structure. Local coordination includes distorted MA$_6$ octahedra and AZ$_4$ tetrahedra. (d) In the 1T$^\prime$ phase, $d$–$p$ band inversion creates spinless Dirac cones with linear dispersion at the $\Lambda$ points along $k_y$. (e) In the SO phase, $d$-$d$ band inversion leads to a quadratic spinless Dirac cone that gaps under SOC, generating quasi-flat bands and van Hove singularities (VHSs) at $\delta_{vhs}$. (f–g) Comparison of spin U(1) symmetry, where the spin axis $\mathbf{S}$ aligns with the $z$-axis, and spin U(1) quasi-symmetry, where $\mathbf{S}$ tilts slightly away from $z$. This tilt leads to deviations from spin Hall quantization in device geometries misaligned with $\mathbf{S}$, and defines the optimal orientation for observing a quantized response.}
\label{fig1}
\end{figure}

\textbf{Lattice and key topological features.}
MoSi$_2$N$_4$ is the first 2D material synthesized without a bulk parent, formed by silicon passivation of a monolayer MoN$_2$~\cite{Hong2020}. It crystallizes in the 1H phase with septuple-layer stacking in the sequence N--Si--N--Mo--N--Si--N, creating a hexagonal lattice with $D_{3h}^1$ symmetry ($P\bar{6}m2$, No.~187) (Fig.~\ref{fig1}(a)). While its symmetry and electronic structure resemble those of 1H-phase transition metal dichalcogenides (TMDs), MoSi$_2$N$_4$ features a thicker, more stable structure with robust electronic properties~\cite{Hong2020,Novesolov2024}. Our earlier work demonstrated that septuple layers in MA$_2$Z$_4$ compounds can be stacked up to the bulk limit, enabling tunable electronic phases via controlling stacking-dependent symmetry~\cite{Rajibul_PRB2021}. A symmetry-lowering structural distortion from the 1H to monoclinic 1T$^\prime$ phase, driven by metal atom dimerization, induces band inversion between transition-metal $d$ and pnictogen $p$ orbitals, leading to a QSH phase~\cite{QSH_Verma2022}. This distortion causes band inversion even without SOC, producing two spinless Dirac cones along the $k_y$ axis (Fig.~\ref{fig1}(c)). Inclusion of SOC hybridizes these cones, resulting in a topologically nontrivial QSH phase with an inverted gap of  $\sim 200$ meV and intrinsic SHC $\sigma_{xy}^z \sim 1.3 e^2/h$.

Structurally, the SO polymorph of MA$_2$Z$_4$ comprises a central MZ$_2$ layer, analogous to that in TMDs, sandwiched between two A--Z layers (Figs.~\ref{fig1}(b)–(c)), with a stacking sequence similar to the 1H phase. The transition-metal atom adopts a distorted MA$_6$ coordination with unequal M–A bond lengths, while the outer layers form AZ$_4$ tetrahedral motifs. This arrangement creates a periodic 2D network of four- and eight-membered rings reminiscent of grain boundary motifs in TMDs~\cite{SoTMD_Exp2013,SOTMD_PRB2015}. Unlike in TMDs, where the SO lattice is unstable and typically relaxes into an orthorhombic phase, MA$_2$Z$_4$ maintains the SO lattice without symmetry breaking~\cite{SOTMD_PRB2015}. Although the SO phase has a higher total energy than the 1H and 1T$^\prime$ polymorphs, phonon dispersion calculations show no imaginary modes throughout the Brillouin zone (see Supplemental Materials (SMs)), confirming dynamical stability. In addition, ab-initio molecular dynamics simulations at 300 K demonstrate that the structure remains intact over time with minimal free energy fluctuations (Fig.~S1 SMs). The SO lattice crystallizes in the $P4/mbm$ (No.~127) space group, exhibiting inversion $\mathcal{I}$, $C_{2z}$, $C_{4z}$, and nonsymmorphic symmetries including $\{C_{2x} | \tfrac{1}{2},\tfrac{1}{2}, 0\}$, $\{C_{2y} | \tfrac{1}{2}, \tfrac{1}{2}, 0\}$, and $\{C_{2xy} | \tfrac{1}{2}, \tfrac{1}{2}, 0\}$. We confirmed the stability of sixteen MA$_2$Z$_4$ compounds with M= Mo, W; A= Si, Ge; and Z= P, N, As, and Sb, and present their optimized structural parameters and topological states in Table~\ref{table1} (see SMs for details). Figures~\ref{fig1}(e)-(g) highlight the key topological electronic and spin spectrum features of SO MA$_2$Z$_4$. Unlike the 1T$^\prime$ phase with $d$–$p$ band inversion (Fig.~\ref{fig1}(d)), the SO phase shows a $d-d$ band inversion yielding quasi-flat bands and VHSs near the Fermi level (Fig.~\ref{fig1}(e)). These materials maintain spin U(1) quasi-symmetry (Figs.~\ref{fig1}(f)–(g)) with a slight tilt from the stacking-normal $z$ axis, producing nearly quantized SHC.

\begin{figure}[t!]
\includegraphics[width=1\linewidth]{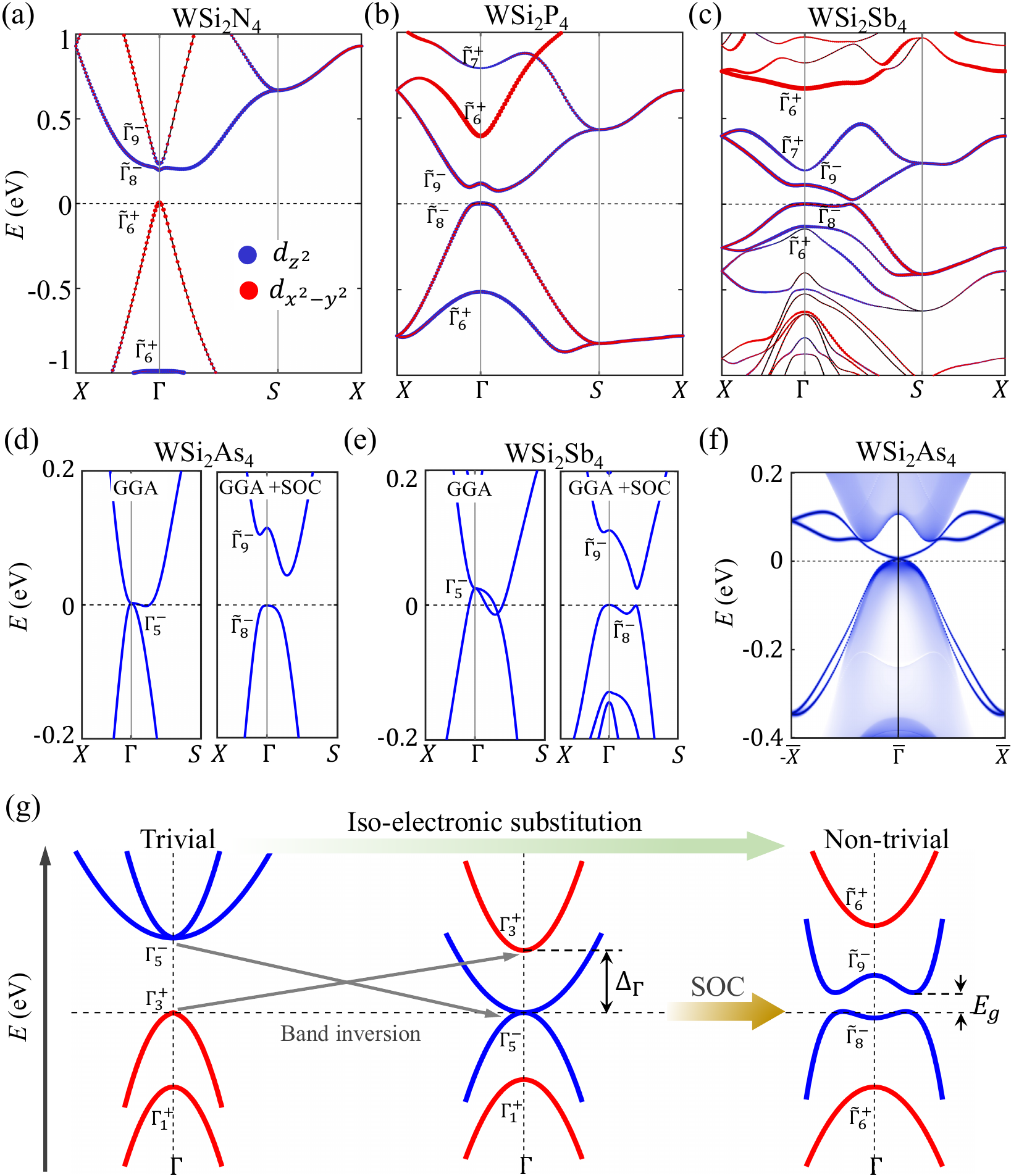}
\caption{\textbf{Band structure and topological characterization}. Orbital-resolved band structure of (a) WSi$_2$N$_4$, (b) WSi$_2$P$_4$, and (c) WSi$_2$Sb$_4$ with spin--orbit coupling (SOC). Red and blue mark transition-metal $d_{x^2-y^2}$ and $d_{z^2}$ orbitals. Irreducible representations (IRs) and associated parities at the $\Gamma$ point are shown. (d)--(e)  Closeup band structures of WSi$_2$As$_4$ and WSi$_2$Sb$_4$ without and with SOC. The doubly degenerate $\Gamma_5^-$ at the Fermi level splits upon inclusion of SOC, opening a gap and driving the system into a QSH phase with $\mathbb{Z}_2=1$. WSi$_2$Sb$_4$ exhibits flatter energy dispersion near the Fermi level with VHSs.  (f) (010) edge spectrum of WSi$_2$As$_4$. (g) Schematic band evolution across the MA$_2$Z$_4$ compounds. Four bands near the Fermi level form a trivial ordering with $\mathbb{Z}_2 = 0$ for $Z=$ N. Substituting heavier pnictogens ($Z=$ P, As, Sb) induces a band inversion between the $\Gamma_3^+$ and $\Gamma_5^-$ states, forming spinless Dirac cones without SOC. Inclusion of SOC opens an inverted gap in $\Gamma_5^-$, yielding a QSH state with $\mathbb{Z}_2 = 1$.} 
\label{bands}
\end{figure}

\textbf{Electronic structures and band inversion.} We present the band structure evolution of monolayers of SO MA$_2$Z$_4$ compounds, using WSi$_2$Z$_4$ as a representative example with varying pnictogen $Z$ atoms in Fig.~\ref{bands}. Orbital-resolved band structures of WSi$_2$N$_4$, WSi$_2$P$_4$, and WSi$_2$Sb$_4$ (Figs.~\ref{bands}(a)-(c)) reveal four low-energy states primarily derived from transition-metal $d_{z^2}$ and $d_{x^2 - y^2}$ orbitals. In the absence of SOC, these states belong to a two-fold degenerate $\Gamma_5^-$ and two singly degenerate $\Gamma_1^+$ and $\Gamma_3^+$ irreducible representations, as schematically shown in Fig.~\ref{bands}(g). Upon including SOC, these states evolve into two $\Gamma_6^+$ bands and one each of $\tilde{\Gamma}_8^-$ and $\tilde{\Gamma}_9^-$, lifting the spinless degeneracies. For WSi$_2$N$_4$, both $\Gamma_6^+$ bands lie below the Fermi level, while $\tilde{\Gamma}_8^-$ and $\tilde{\Gamma}_9^-$ remain above, resulting in a trivial insulating state. Replacing N with P induces a band inversion between $\Gamma_5^-$ and $\Gamma_3^+$ without SOC, forming a parabolic crossing at the Fermi level characteristic of a spinless Dirac node. As these states have the same parity, SOC opens a gap between the resulting $\tilde{\Gamma}_8^-$ and $\tilde{\Gamma}_9^-$ bands, driving the system into a topologically nontrivial insulating phase. Introducing heavier pnictogens (As, Sb) further increases the inverted gap and flattens the bands near the Fermi level (Figs.~\ref{bands}(d)-(e)). In WSi$_2$Sb$_4$, additional band warping along the $\Gamma$–S direction generates VHSs that coexist with the inverted band structure. This trend persists across the MA$_2$Z$_4$ family: for Z = N, the system remains topologically trivial, while Z =  P, As, or Sb drives a $d-d$ band inversion and induces a nontrivial topological phase.

To confirm the nontrivial topology, we calculate the $\mathbb{Z}_2$ invariant from parity eigenvalues at time-reversal-invariant momenta (TRIM) and analyze the edge-state spectrum~\cite{Fu2007}. The parity products of occupied bands at $\Gamma$, $X/Y$, and $S$ are $-$, $+/+$, and $+$, respectively, resulting in $\mathbb{Z}_2 = 1$ for MA$_2$Z$_4$ with Z = P, As, or Sb, and $\mathbb{Z}_2 = 0$ for Z = N. The (010) edge spectrum of WSi$_2$As$_4$ (Fig.~\ref{bands}(f)) reveals a pair of topological edge states traversing the $\overline{\Gamma}$–$\overline{X}$ path inside the bulk gap, connecting valence and conduction bands. These edge-state dispersions vary with atomic composition, providing a means to tune the topological properties within MA$_2$Z$_4$ compounds. The overall band inversion mechanism and topological classifications of MA$_2$Z$_4$ are summarized in Fig.~\ref{bands}(g) and Table~\ref{table1}.

\renewcommand{\arraystretch}{1.25}
\begin{table*}[ht!]
\caption{Calculated structural and electronic properties of monolayer SO MA$_2$Z$_4$ compounds (M = Mo, W; A = Si, Ge; $Z =$ pnictogen). The in-plane lattice constants $a = b$, global band gap $E_g$, and inverted band gap $\Delta_{\Gamma}= E_{\Gamma_5^-} - E_{\Gamma_3^+}$ at the $\Gamma$ point are listed. $\mathbb{Z}_2$ topological invariant, resulting topological phase, and SHC $\sigma_{xy}^z$ are also provided. Trivial Ins. and QSH denote trivial and quantum spin Hall insulator, respectively.}
\centering
\begin{tabularx}{\textwidth}{Y Y Y Y Y Y Y} 
\hline\hline
Material & $a=b$ (\AA) & Inverted gap (meV) [$\Delta_\Gamma$] & Band gap (meV) [$E_g$] & Topological invariant [$ \mathbb{Z}_2]$ & Topological state & SHC ($e^2/h$) \\
\hline
MoSi$_2$N$_4$ & 5.694 & 299    & 281  & 0 & Trivial Ins. & 0 \\
WSi$_2$N$_4$  & 5.701 & 241    & 191  & 0 & Trivial Ins. & 0 \\
MoGe$_2$N$_4$ & 5.928 & 92   & 74 & 0 & Trivial Ins. & 0 \\
WGe$_2$N$_4$  & 5.931 & 101    & 34   & 0 & Trivial Ins. & 0 \\
MoSi$_2$P$_4$ & 6.831 & -42    & 22   & 1 & QSH & 1.99 \\
WSi$_2$P$_4$  & 6.851 & -269     & 74   & 1 & QSH & 1.96 \\
MoGe$_2$P$_4$ & 7.010 & -27      & 18   & 1 & QSH & 1.99 \\
WGe$_2$P$_4$  & 7.024 & -238     & 60   & 1 & QSH & 1.99 \\
MoSi$_2$As$_4$& 7.130 & -193     & 28   & 1 & QSH & 1.98 \\
WSi$_2$As$_4$ & 7.149 & -431     & 45   & 1 & QSH & 1.91 \\
MoGe$_2$As$_4$& 7.299 & -156     & 25   & 1 & QSH & 1.99 \\
WGe$_2$As$_4$ & 7.309 & -392     & -20 & 1 & QSH& 1.95 \\
MoSi$_2$Sb$_4$& 7.681 & -258     & 29   & 1 & QSH & 1.97 \\
WSi$_2$Sb$_4$ & 7.701 & -562     & 24   & 1 & QSH & 1.79 \\
MoGe$_2$Sb$_4$& 7.816 & -206     & 0  & - & Metal  & - \\
WGe$_2$Sb$_4$ & 7.831 & -533     & 0  & - & Metal & - \\
\hline\hline
\end{tabularx}
\label{table1}
\end{table*}

\textbf{Feature spectrum and quantum spin Hall quantization.} In an ideal 2D QSH insulator with spin rotational symmetry about the $z$-axis (spin U(1) symmetry), only the $z$-component of the SHC, $\sigma_{xy}^z$, is symmetry-allowed~\cite{QSH_KM2005}. (Here, $\sigma_{xy}^z$ represents the $z$-polarized spin current along the $x$ direction generated by an electric field applied along the $y$ direction.) This constraint follows from Neumann’s principle, which states that any physical property of a crystal must respect its point group symmetries and thus constrains allowed physical responses. In realistic materials, however, factors such as finite thickness, strong SOC, and crystal symmetries often break exact $S_z$ conservation, causing partial spin mixing and deviations from quantized $\sigma_{xy}^z$. Despite this, Bloch states remain approximate eigenstates of $S_z$, and the extent of deviation from exact $S_z$ eigenvalues characterizes a spin U(1) quasi-symmetry. To analyze this in MA$_2$Z$_4$ compounds, we calculate spin-feature spectrum~\cite{SCN_PRB2009,FST_Wang2023,PhysRevB.109.155143} by constructing a feature operator $F = P \hat{S}_z P$, which projects the spin operator $\hat{S}_z$ onto occupied Bloch states. When $S_z$ is conserved, the feature spectrum consists of two flat bands at eigenvalues $\pm \hbar/2$ separated by a spin gap $\Delta_{S_z} = \hbar$. With spin mixing, these eigenvalues deviate, but as long as the spin gap remains close to $\hbar$, the system possesses a spin U(1) quasi-symmetry. This framework enables extracting topological invariants from the Wilson loop computed within each spin sector and determining the degree of deviation from exact quantization of $\sigma_{xy}^z$.

\begin{figure}[ht!]
\includegraphics[width=1\linewidth]{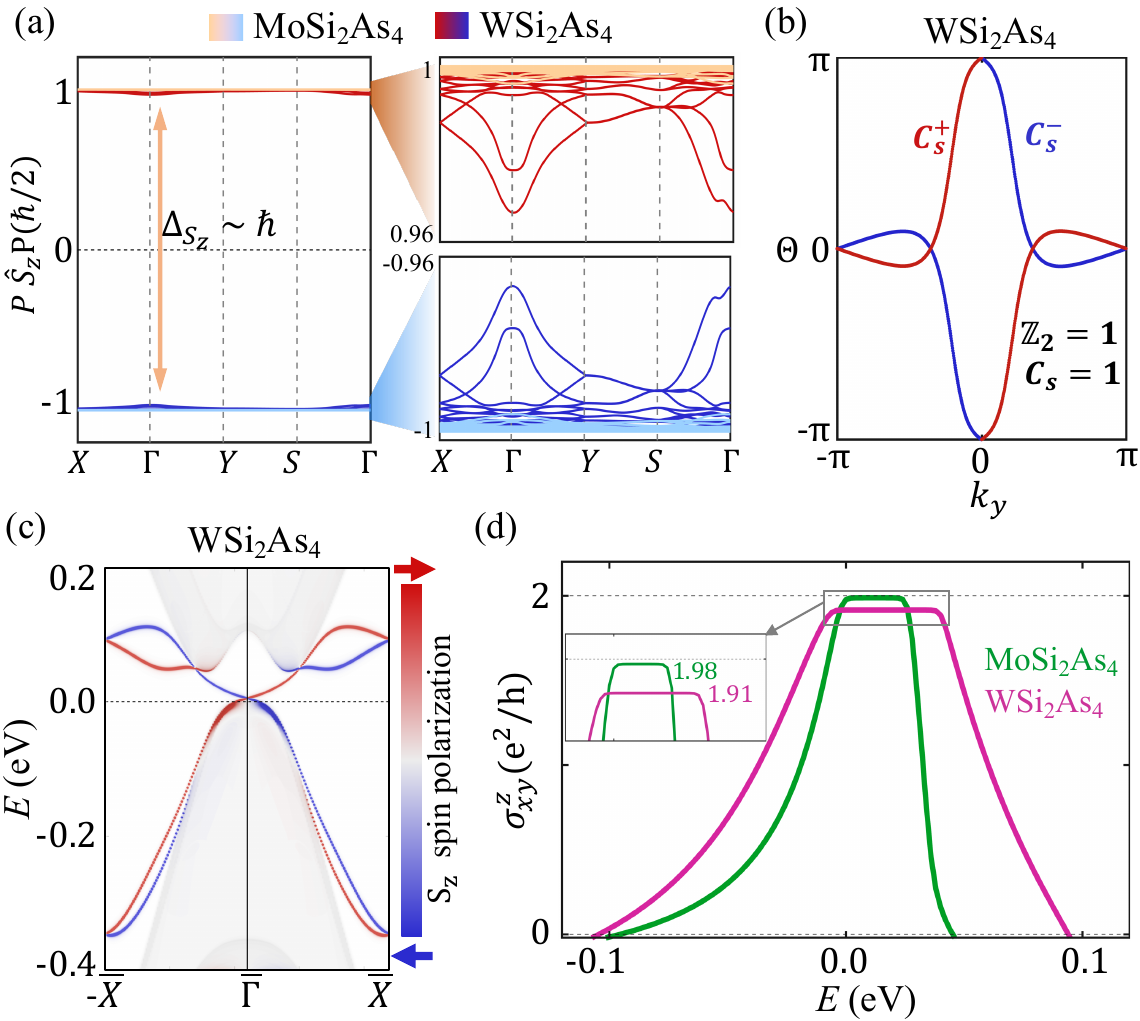}
\caption{\textbf{Feature spectrum topology and spin U(1) quasi-symmetry}. (a) Spin feature spectrum ($P\hat{S}_z P$) of the occupied bulk states of MoSi$_2$As$_4$ and WSi$_2$As$_4$. Eigenvalues remain nearly flat and pinned at $\pm \hbar/2$, with slight deviations near the $\Gamma$ point. The zoomed view shows a larger deviation in WSi$_2$As$_4$, indicating stronger spin-axis tilt from the $z$ direction. The spin gap $\Delta_{S_z} \sim \hbar$ reflects an spin U(1) quasi-symmetry. (b) Spin-resolved Wilson loop spectrum of WSi$_2$As$_4$ showing Chern numbers $+1$ and $-1$ for spin-up and spin-down sectors, giving a total spin Chern number $C_S = 1$. (c) (010) edge state spin texture of WSi$_2$As$_4$ showing counter-propagating $S_z$-polarized edge states. (d) Intrinsic SHC $\sigma_{xy}^z$ of MoSi$_2$As$_4$ and WSi$_2$As$_4$ with a nearly quantized plateau of $\sim 2e^2/h$ within the bulk gap.}
\label{spintop}
\end{figure}

Figure~\ref{spintop}(a) shows the calculated $P \hat{S}_z P$ spin-feature spectrum of the occupied states for MoSi$_2$As$_4$ and WSi$_2$As$_4$, with two distinct sectors corresponding to spin-up and spin-down channels. The $P \hat{S}_z P$ eigenvalues remain nearly flat at $\pm \hbar/2$, with slight deviations near the $\Gamma$ point. The finite spin gap $\Delta_{S_z} \sim \hbar$ indicates approximate spin-rotational symmetry about the $z$-axis consistent with a spin U(1) quasi-symmetry. MoSi$_2$As$_4$ stays close to the ideal case, while WSi$_2$As$_4$ shows larger deviations (eigenvalues spread close to $\pm 0.96 \hbar/2$), implying a stronger tilt of the spin axis away from $z$. The gapped $S_z$ spectrum allows for the computation of the $\hat{S}_z$-resolved Wilson loop in Fig.~\ref{spintop}(b), which yields spin-resolved Chern numbers $C_s^+ = +1$ and $C_s^- = -1$ based on the opposite windings of the Wannier charge centers. This results in a total nonzero spin-Chern number $C_s = (C_s^+ - C_s^-)/2 = 1$. The corresponding (010) edge spectrum in Fig.~\ref{spintop}(c) reveals counter-propagating helical edge states with dominating $S_z$ polarization and a Dirac point at $\overline{\Gamma}$. This is consistent with the nontrivial spin Chern number $C_s$ and the spin-feature spectrum. In contrast, the spectra for $P \hat{S_x}P$ and $P \hat{S_y} P$ remain gapless (see SMs), with $\Delta_{S_x} \sim 0$ and $\Delta_{S_y} \sim 0$, indicating strong mixing of these spin components that further support predominant spin alignment along $z$, as illustrated in Fig.~\ref{fig1}(g).

The observed near $S_z$ polarization in the spin spectrum indicates an almost quantized SHC. To confirm this, we explicitly calculate the intrinsic SHC$\sigma_{xy}^z$ using the Kubo formula~\cite{SHC_TaAs,SHC_w90}. 
\begin{equation}
\sigma_{xy}^{z} = \frac{-e^2}{\hbar} \frac{1}{A_{2D}} \sum_{\bm{k}} \Omega_{xy}^z (\bm{k}),
\end{equation}
where
\begin{equation}
\Omega_{xy}^z(\bm{k}) = \sum_n f_{n} (\bm{k}) \, \Omega_{n,xy}^z(\bm{k})
\end{equation}
is the $k$-resolved spin Berry curvature, and
\begin{equation}
\Omega_{n,xy}^z(\bm{k}) = \hbar^2 \sum_{m \ne n} \frac{-2\,\text{Im}\big[\langle n\bm{k}| \hat{J}_x^z | m\bm{k}\rangle \langle m\bm{k}| \hat{v}_y | n\bm{k}\rangle \big]}{(E_{n\bm{k}} - E_{m\bm{k}})^2}
\end{equation}
gives the band-resolved spin Berry curvature in the 2D Brillouin zone with area $A_{2D}$. Here, $E_{n\bm{k}}$ denotes the energy of the Bloch state $|n\bm{k}\rangle$ with occupation $f_{n}(\bm{k})$. The spin current operator is defined as $\hat{J}_x^z = \tfrac{1}{2}\{\hat{\sigma}_z, \hat{v}_x\}$, where $\hat{\sigma}_z$ is the spin operator and $\hat{v}_x$ is the velocity operator.  We compute the SHC of all SO monolayers using a dense $k$-grid of $10^5$ points in the 2D Brillouin zone. Figure~\ref{spintop}(d) shows the variation of $\sigma_{xy}^z$ with energy for WSi$_2$As$_4$ and MoSi$_2$As$_4$, revealing a nearly quantized plateau close to $2e^2/h$ within the band gap. The magnitude of $\sigma_{xy}^z$ in MoSi$_2$As$_4$ reaches about $1.98 e^2/h$, slightly higher than $1.91 e^2/h$ in WSi$_2$As$_4$. These values agree with the $P \hat{S}_z P$ spin-feature spectrum and confirm the spin U(1) quasi-symmetry. All SO compounds in the QSH phase show similarly nearly quantized $\sigma_{xy}^z$, with plateau magnitudes summarized in Table~\ref{table1}.

 \begin{figure}[ht!]
\includegraphics[width=1\linewidth]{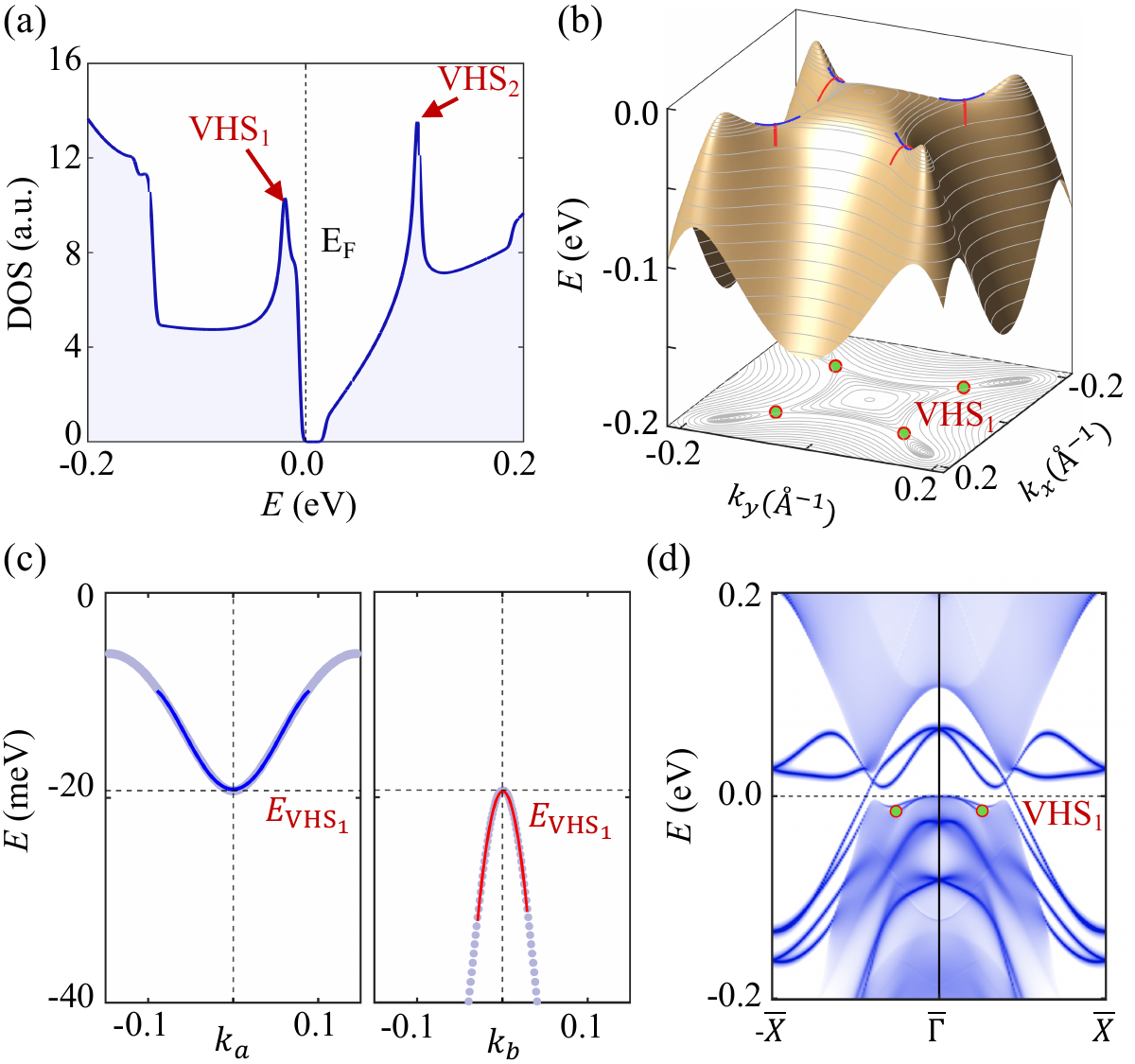}
\caption{\textbf{High-order VHSs in WSi$_2$Sb$_4$}. (a) Density of states (DOS) showing two sharp peaks associated with VHSs (VHS$_1$ and VHS$_2$) at $-19$ meV and $102$ meV near the Fermi level. (b) Valence band energy dispersion reveals quasi-flat regions with four symmetry-related saddle-point VHSs at generic $k$ points along the $\Gamma-S$ directions. The constant energy contour marks these VHSs locations. (c) Local dispersion near VHS$_1$ along principal axes $k_a$ and $k_b$, with polynomial fits: fourth-order along $k_a$ (blue) and second-order along $k_b$ (red), indicating higher-order VHSs. (d) (010) edge spectrum showing coexistence of VHSs and topological edge states.}
\label{hovhs}
\end{figure}

\textbf{High-order VHSs.} Due to $d-d$ band inversion and quadratic band touchings without SOC, all SO monolayers display quasi-flat bands near the Fermi level, with heavier transition metal compounds showing flatter dispersions. The MA$_2$(As, Sb)$_4$ band structures exhibit quasi-flat bands near the Fermi level (see SMs). Using WSi$_2$Sb$_4$ as an example, we highlight quasi-flat bands and VHSs accompanying the QSH state in Fig.~\ref{hovhs}. The density of states (Fig.~\ref{hovhs}(a)) shows two peaks at $-19$ meV (VHS$_1$) and $102$ meV (VHS$_2$), linked to VHSs in the inverted valence and conduction bands. Figure~\ref{hovhs}(b) presents the valence band dispersion with a clear flat region and four symmetry-related VHSs (VHS$_1$) along the $\Gamma$–$S$ direction. These VHSs occur at $(k_x, k_y) = \left(0.10, 0.10 \right)$ \AA$^{-1}$ at $-19$ meV. The constant energy contours near these points cross tangentially, indicating high-order VHSs, unlike conventional 2D VHSs with linear crossings~\cite{Yuan2019,Kagome_Patra2025}. Polynomial fits of VHS$_1$ dispersion along principal axes ($k_a$, $k_b$) in Fig.~\ref{hovhs}(c) reveal quartic dependence in $k_a$ and quadratic in $k_b$ described by $E = -100 k_a^4 + 2 k_a^2 - 14 k_b^2 - 0.019$. This anisotropy matches the asymmetric DOS near VHS$_1$ in Fig.~\ref{hovhs}(a), showing a power-law divergence characteristic of high-order VHSs~\cite{Yuan2019}. The (010) edge spectrum of WSi$_2$Sb$_4$ in Fig.~\ref{hovhs}(d) shows high-order VHSs coexisting with topological edge states. These VHSs and flat bands stay robust under various exchange-correlation functionals and lie close to the Fermi level (see SMs), reachable with slight hole doping. The combination of quasi-flat bands and high-order VHSs within the QSH phase makes WSi$_2$Sb$_4$ a strong candidate for correlation-driven instabilities and dual topological phases.

\textbf{Summary.} 
We identify a new SO polymorph in the MA$_2$Z$_4$ materials family as a promising platform for QSH insulators that exhibit nearly quantized SHC and high-order VHSs. These materials feature four bands near the Fermi level derived from transition metal $d$ orbitals forming a SO lattice with Pnictogen atoms. A $d-d$ band inversion among these states produces a nontrivial $\mathbb{Z}_2=1$ invariant and spin Chern number $C_S = 1$. Spin-feature spectrum analysis reveals an approximately $S_z$-polarized Hamiltonian with spin U(1) quasi-symmetry, resulting in SHC close to $2e^2/h$ within the inverted gap. Additionally, MA$_2$(As, Sb)$_4$ compounds exhibit quasi-flat bands, with WSi$_2$Sb$_4$ hosting multiple high-order VHSs at generic momenta near the Fermi level.
Beyond identifying these new QSH insulators featuring VHSs, our results demonstrate that the spin-feature spectrum method provides an effective tool to quantify deviations from exact $S_z$ conservation and to detect topological phases with nearly quantized SHC. The coexistence of high-order VHSs and QSH topology in these materials offers a unique platform to explore correlation-driven phenomena. Considering that TaIrTe$_4$ remains the only experimentally confirmed density-tuned correlated topological insulator, these SO lattice materials could emerge as promising candidates to realize intertwined QSH and correlated phases.

\section*{Acknowledgements}
We thank Yueh-Ting Yao and Tay-Rong Chang for valuable discussion. This work is supported by the Department of Atomic Energy of the Government of India under Project No. 12-R$\&$D-TFR-5.10-0100 and benefited from the computational resources of TIFR Mumbai.
\bibliography{sqoct}

\begin{thebibliography}{56}%
\makeatletter
\providecommand \@ifxundefined [1]{%
 \@ifx{#1\undefined}
}%
\providecommand \@ifnum [1]{%
 \ifnum #1\expandafter \@firstoftwo
 \else \expandafter \@secondoftwo
 \fi
}%
\providecommand \@ifx [1]{%
 \ifx #1\expandafter \@firstoftwo
 \else \expandafter \@secondoftwo
 \fi
}%
\providecommand \natexlab [1]{#1}%
\providecommand \enquote  [1]{``#1''}%
\providecommand \bibnamefont  [1]{#1}%
\providecommand \bibfnamefont [1]{#1}%
\providecommand \citenamefont [1]{#1}%
\providecommand \href@noop [0]{\@secondoftwo}%
\providecommand \href [0]{\begingroup \@sanitize@url \@href}%
\providecommand \@href[1]{\@@startlink{#1}\@@href}%
\providecommand \@@href[1]{\endgroup#1\@@endlink}%
\providecommand \@sanitize@url [0]{\catcode `\\12\catcode `\$12\catcode
  `\&12\catcode `\#12\catcode `\^12\catcode `\_12\catcode `\%12\relax}%
\providecommand \@@startlink[1]{}%
\providecommand \@@endlink[0]{}%
\providecommand \url  [0]{\begingroup\@sanitize@url \@url }%
\providecommand \@url [1]{\endgroup\@href {#1}{\urlprefix }}%
\providecommand \urlprefix  [0]{URL }%
\providecommand \Eprint [0]{\href }%
\providecommand \doibase [0]{https://doi.org/}%
\providecommand \selectlanguage [0]{\@gobble}%
\providecommand \bibinfo  [0]{\@secondoftwo}%
\providecommand \bibfield  [0]{\@secondoftwo}%
\providecommand \translation [1]{[#1]}%
\providecommand \BibitemOpen [0]{}%
\providecommand \bibitemStop [0]{}%
\providecommand \bibitemNoStop [0]{.\EOS\space}%
\providecommand \EOS [0]{\spacefactor3000\relax}%
\providecommand \BibitemShut  [1]{\csname bibitem#1\endcsname}%
\let\auto@bib@innerbib\@empty
\bibitem [{\citenamefont {Singh}\ \emph {et~al.}(2023)\citenamefont {Singh},
  \citenamefont {Lin},\ and\ \citenamefont {Bansil}}]{AM_Singh2022}%
  \BibitemOpen
  \bibfield  {author} {\bibinfo {author} {\bibfnamefont {B.}~\bibnamefont
  {Singh}}, \bibinfo {author} {\bibfnamefont {H.}~\bibnamefont {Lin}},\ and\
  \bibinfo {author} {\bibfnamefont {A.}~\bibnamefont {Bansil}},\ }\bibfield
  {title} {\bibinfo {title} {Topology and symmetry in quantum materials},\
  }\href {https://doi.org/10.1002/adma.202201058} {\bibfield  {journal}
  {\bibinfo  {journal} {Adv. Mater.}\ }\textbf {\bibinfo {volume} {35}},\
  \bibinfo {pages} {2201058} (\bibinfo {year} {2023})}\BibitemShut {NoStop}%
\bibitem [{\citenamefont {Hasan}\ and\ \citenamefont
  {Kane}(2010)}]{RMP_Hasan2010}%
  \BibitemOpen
  \bibfield  {author} {\bibinfo {author} {\bibfnamefont {M.~Z.}\ \bibnamefont
  {Hasan}}\ and\ \bibinfo {author} {\bibfnamefont {C.~L.}\ \bibnamefont
  {Kane}},\ }\bibfield  {title} {\bibinfo {title} {Colloquium: Topological
  insulators},\ }\href {https://doi.org/10.1103/RevModPhys.82.3045} {\bibfield
  {journal} {\bibinfo  {journal} {Rev. Mod. Phys.}\ }\textbf {\bibinfo {volume}
  {82}},\ \bibinfo {pages} {3045} (\bibinfo {year} {2010})}\BibitemShut
  {NoStop}%
\bibitem [{\citenamefont {Bansil}\ \emph {et~al.}(2016)\citenamefont {Bansil},
  \citenamefont {Lin},\ and\ \citenamefont {Das}}]{RMP_Bansil2016}%
  \BibitemOpen
  \bibfield  {author} {\bibinfo {author} {\bibfnamefont {A.}~\bibnamefont
  {Bansil}}, \bibinfo {author} {\bibfnamefont {H.}~\bibnamefont {Lin}},\ and\
  \bibinfo {author} {\bibfnamefont {T.}~\bibnamefont {Das}},\ }\bibfield
  {title} {\bibinfo {title} {Colloquium: Topological band theory},\ }\href
  {https://doi.org/10.1103/RevModPhys.88.021004} {\bibfield  {journal}
  {\bibinfo  {journal} {Rev. Mod. Phys.}\ }\textbf {\bibinfo {volume} {88}},\
  \bibinfo {pages} {021004} (\bibinfo {year} {2016})}\BibitemShut {NoStop}%
\bibitem [{\citenamefont {Hasan}\ \emph {et~al.}(2021)\citenamefont {Hasan},
  \citenamefont {Chang}, \citenamefont {Belopolski}, \citenamefont {Bian},
  \citenamefont {Xu},\ and\ \citenamefont {Yin}}]{NR_Hasan2021}%
  \BibitemOpen
  \bibfield  {author} {\bibinfo {author} {\bibfnamefont {M.~Z.}\ \bibnamefont
  {Hasan}}, \bibinfo {author} {\bibfnamefont {G.}~\bibnamefont {Chang}},
  \bibinfo {author} {\bibfnamefont {I.}~\bibnamefont {Belopolski}}, \bibinfo
  {author} {\bibfnamefont {G.}~\bibnamefont {Bian}}, \bibinfo {author}
  {\bibfnamefont {S.-Y.}\ \bibnamefont {Xu}},\ and\ \bibinfo {author}
  {\bibfnamefont {J.-X.}\ \bibnamefont {Yin}},\ }\bibfield  {title} {\bibinfo
  {title} {Weyl, {D}irac and high-fold chiral fermions in topological quantum
  matter},\ }\href {https://doi.org/10.1038/s41578-021-00301-3} {\bibfield
  {journal} {\bibinfo  {journal} {Nat. Rev. Mater.}\ }\textbf {\bibinfo
  {volume} {6}},\ \bibinfo {pages} {784–803} (\bibinfo {year}
  {2021})}\BibitemShut {NoStop}%
\bibitem [{\citenamefont {Xiao}\ and\ \citenamefont {Yan}(2021)}]{NR_Xiao2021}%
  \BibitemOpen
  \bibfield  {author} {\bibinfo {author} {\bibfnamefont {J.}~\bibnamefont
  {Xiao}}\ and\ \bibinfo {author} {\bibfnamefont {B.}~\bibnamefont {Yan}},\
  }\bibfield  {title} {\bibinfo {title} {First-principles calculations for
  topological quantum materials},\ }\href
  {https://doi.org/10.1038/s42254-021-00292-8} {\bibfield  {journal} {\bibinfo
  {journal} {Nat. Rev. Phys.}\ }\textbf {\bibinfo {volume} {3}},\ \bibinfo
  {pages} {283–297} (\bibinfo {year} {2021})}\BibitemShut {NoStop}%
\bibitem [{\citenamefont {Tokura}\ \emph {et~al.}(2017)\citenamefont {Tokura},
  \citenamefont {Kawasaki},\ and\ \citenamefont {Nagaosa}}]{Tokura2017}%
  \BibitemOpen
  \bibfield  {author} {\bibinfo {author} {\bibfnamefont {Y.}~\bibnamefont
  {Tokura}}, \bibinfo {author} {\bibfnamefont {M.}~\bibnamefont {Kawasaki}},\
  and\ \bibinfo {author} {\bibfnamefont {N.}~\bibnamefont {Nagaosa}},\
  }\bibfield  {title} {\bibinfo {title} {Emergent functions of quantum
  materials},\ }\href {https://doi.org/10.1038/nphys4274} {\bibfield  {journal}
  {\bibinfo  {journal} {Nat. Phys.}\ }\textbf {\bibinfo {volume} {13}},\
  \bibinfo {pages} {1056–1068} (\bibinfo {year} {2017})}\BibitemShut
  {NoStop}%
\bibitem [{\citenamefont {Kane}\ and\ \citenamefont
  {Mele}(2005{\natexlab{a}})}]{QSH_KM2005}%
  \BibitemOpen
  \bibfield  {author} {\bibinfo {author} {\bibfnamefont {C.~L.}\ \bibnamefont
  {Kane}}\ and\ \bibinfo {author} {\bibfnamefont {E.~J.}\ \bibnamefont
  {Mele}},\ }\bibfield  {title} {\bibinfo {title} {Quantum spin hall effect in
  graphene},\ }\href {https://doi.org/10.1103/PhysRevLett.95.226801} {\bibfield
   {journal} {\bibinfo  {journal} {Phys. Rev. Lett.}\ }\textbf {\bibinfo
  {volume} {95}},\ \bibinfo {pages} {226801} (\bibinfo {year}
  {2005}{\natexlab{a}})}\BibitemShut {NoStop}%
\bibitem [{\citenamefont {Kane}\ and\ \citenamefont
  {Mele}(2005{\natexlab{b}})}]{QSH_KM2005b}%
  \BibitemOpen
  \bibfield  {author} {\bibinfo {author} {\bibfnamefont {C.~L.}\ \bibnamefont
  {Kane}}\ and\ \bibinfo {author} {\bibfnamefont {E.~J.}\ \bibnamefont
  {Mele}},\ }\bibfield  {title} {\bibinfo {title} {${Z}_{2}$ topological order
  and the quantum spin hall effect},\ }\href
  {https://doi.org/10.1103/PhysRevLett.95.146802} {\bibfield  {journal}
  {\bibinfo  {journal} {Phys. Rev. Lett.}\ }\textbf {\bibinfo {volume} {95}},\
  \bibinfo {pages} {146802} (\bibinfo {year} {2005}{\natexlab{b}})}\BibitemShut
  {NoStop}%
\bibitem [{\citenamefont {Bernevig}\ \emph {et~al.}(2006)\citenamefont
  {Bernevig}, \citenamefont {Hughes},\ and\ \citenamefont
  {Zhang}}]{HgTe_Bernevig2006}%
  \BibitemOpen
  \bibfield  {author} {\bibinfo {author} {\bibfnamefont {B.~A.}\ \bibnamefont
  {Bernevig}}, \bibinfo {author} {\bibfnamefont {T.~L.}\ \bibnamefont
  {Hughes}},\ and\ \bibinfo {author} {\bibfnamefont {S.-C.}\ \bibnamefont
  {Zhang}},\ }\bibfield  {title} {\bibinfo {title} {Quantum spin hall effect
  and topological phase transition in {H}g{T}e quantum wells},\ }\href
  {https://doi.org/10.1126/science.1133734} {\bibfield  {journal} {\bibinfo
  {journal} {Science}\ }\textbf {\bibinfo {volume} {314}},\ \bibinfo {pages}
  {1757–1761} (\bibinfo {year} {2006})}\BibitemShut {NoStop}%
\bibitem [{\citenamefont {Konig}\ \emph {et~al.}(2007)\citenamefont {Konig},
  \citenamefont {Wiedmann}, \citenamefont {Brune}, \citenamefont {Roth},
  \citenamefont {Buhmann}, \citenamefont {Molenkamp}, \citenamefont {Qi},\ and\
  \citenamefont {Zhang}}]{HgTe_Konig2007}%
  \BibitemOpen
  \bibfield  {author} {\bibinfo {author} {\bibfnamefont {M.}~\bibnamefont
  {Konig}}, \bibinfo {author} {\bibfnamefont {S.}~\bibnamefont {Wiedmann}},
  \bibinfo {author} {\bibfnamefont {C.}~\bibnamefont {Brune}}, \bibinfo
  {author} {\bibfnamefont {A.}~\bibnamefont {Roth}}, \bibinfo {author}
  {\bibfnamefont {H.}~\bibnamefont {Buhmann}}, \bibinfo {author} {\bibfnamefont
  {L.~W.}\ \bibnamefont {Molenkamp}}, \bibinfo {author} {\bibfnamefont {X.-L.}\
  \bibnamefont {Qi}},\ and\ \bibinfo {author} {\bibfnamefont {S.-C.}\
  \bibnamefont {Zhang}},\ }\bibfield  {title} {\bibinfo {title} {Quantum spin
  hall insulator state in {H}g{T}e quantum wells},\ }\href
  {https://doi.org/10.1126/science.1148047} {\bibfield  {journal} {\bibinfo
  {journal} {Science}\ }\textbf {\bibinfo {volume} {318}},\ \bibinfo {pages}
  {766–770} (\bibinfo {year} {2007})}\BibitemShut {NoStop}%
\bibitem [{\citenamefont {Prodan}(2009)}]{SCN_PRB2009}%
  \BibitemOpen
  \bibfield  {author} {\bibinfo {author} {\bibfnamefont {E.}~\bibnamefont
  {Prodan}},\ }\bibfield  {title} {\bibinfo {title} {Robustness of the
  spin-chern number},\ }\href {https://doi.org/10.1103/PhysRevB.80.125327}
  {\bibfield  {journal} {\bibinfo  {journal} {Phys. Rev. B}\ }\textbf {\bibinfo
  {volume} {80}},\ \bibinfo {pages} {125327} (\bibinfo {year}
  {2009})}\BibitemShut {NoStop}%
\bibitem [{\citenamefont {Matusalem}\ \emph {et~al.}(2019)\citenamefont
  {Matusalem}, \citenamefont {Marques}, \citenamefont {Teles}, \citenamefont
  {Matthes}, \citenamefont {Furthm\"uller},\ and\ \citenamefont
  {Bechstedt}}]{SHC_PRB2019}%
  \BibitemOpen
  \bibfield  {author} {\bibinfo {author} {\bibfnamefont {F.}~\bibnamefont
  {Matusalem}}, \bibinfo {author} {\bibfnamefont {M.}~\bibnamefont {Marques}},
  \bibinfo {author} {\bibfnamefont {L.~K.}\ \bibnamefont {Teles}}, \bibinfo
  {author} {\bibfnamefont {L.}~\bibnamefont {Matthes}}, \bibinfo {author}
  {\bibfnamefont {J.}~\bibnamefont {Furthm\"uller}},\ and\ \bibinfo {author}
  {\bibfnamefont {F.}~\bibnamefont {Bechstedt}},\ }\bibfield  {title} {\bibinfo
  {title} {Quantization of spin hall conductivity in two-dimensional
  topological insulators versus symmetry and spin-orbit interaction},\ }\href
  {https://doi.org/10.1103/PhysRevB.100.245430} {\bibfield  {journal} {\bibinfo
   {journal} {Phys. Rev. B}\ }\textbf {\bibinfo {volume} {100}},\ \bibinfo
  {pages} {245430} (\bibinfo {year} {2019})}\BibitemShut {NoStop}%
\bibitem [{\citenamefont {Monaco}\ and\ \citenamefont
  {Ul\ifmmode~\check{c}\else \v{c}\fi{}akar}(2020)}]{SHC_PRB2020}%
  \BibitemOpen
  \bibfield  {author} {\bibinfo {author} {\bibfnamefont {D.}~\bibnamefont
  {Monaco}}\ and\ \bibinfo {author} {\bibfnamefont {L.}~\bibnamefont
  {Ul\ifmmode~\check{c}\else \v{c}\fi{}akar}},\ }\bibfield  {title} {\bibinfo
  {title} {Spin hall conductivity in insulators with nonconserved spin},\
  }\href {https://doi.org/10.1103/PhysRevB.102.125138} {\bibfield  {journal}
  {\bibinfo  {journal} {Phys. Rev. B}\ }\textbf {\bibinfo {volume} {102}},\
  \bibinfo {pages} {125138} (\bibinfo {year} {2020})}\BibitemShut {NoStop}%
\bibitem [{\citenamefont {Zhao}\ \emph {et~al.}(2021)\citenamefont {Zhao},
  \citenamefont {Runburg}, \citenamefont {Fei}, \citenamefont {Mutch},
  \citenamefont {Malinowski}, \citenamefont {Sun}, \citenamefont {Huang},
  \citenamefont {Pesin}, \citenamefont {Cui}, \citenamefont {Xu}, \citenamefont
  {Chu},\ and\ \citenamefont {Cobden}}]{SC_PRX2021}%
  \BibitemOpen
  \bibfield  {author} {\bibinfo {author} {\bibfnamefont {W.}~\bibnamefont
  {Zhao}}, \bibinfo {author} {\bibfnamefont {E.}~\bibnamefont {Runburg}},
  \bibinfo {author} {\bibfnamefont {Z.}~\bibnamefont {Fei}}, \bibinfo {author}
  {\bibfnamefont {J.}~\bibnamefont {Mutch}}, \bibinfo {author} {\bibfnamefont
  {P.}~\bibnamefont {Malinowski}}, \bibinfo {author} {\bibfnamefont
  {B.}~\bibnamefont {Sun}}, \bibinfo {author} {\bibfnamefont {X.}~\bibnamefont
  {Huang}}, \bibinfo {author} {\bibfnamefont {D.}~\bibnamefont {Pesin}},
  \bibinfo {author} {\bibfnamefont {Y.-T.}\ \bibnamefont {Cui}}, \bibinfo
  {author} {\bibfnamefont {X.}~\bibnamefont {Xu}}, \bibinfo {author}
  {\bibfnamefont {J.-H.}\ \bibnamefont {Chu}},\ and\ \bibinfo {author}
  {\bibfnamefont {D.~H.}\ \bibnamefont {Cobden}},\ }\bibfield  {title}
  {\bibinfo {title} {Determination of the spin axis in quantum spin hall
  insulator candidate monolayer {WT}e$_{2}$},\ }\href
  {https://doi.org/10.1103/PhysRevX.11.041034} {\bibfield  {journal} {\bibinfo
  {journal} {Phys. Rev. X}\ }\textbf {\bibinfo {volume} {11}},\ \bibinfo
  {pages} {041034} (\bibinfo {year} {2021})}\BibitemShut {NoStop}%
\bibitem [{\citenamefont {Garcia}\ \emph {et~al.}(2020)\citenamefont {Garcia},
  \citenamefont {Vila}, \citenamefont {Hsu}, \citenamefont {Waintal},
  \citenamefont {Pereira},\ and\ \citenamefont {Roche}}]{SC_PRL2020}%
  \BibitemOpen
  \bibfield  {author} {\bibinfo {author} {\bibfnamefont {J.~H.}\ \bibnamefont
  {Garcia}}, \bibinfo {author} {\bibfnamefont {M.}~\bibnamefont {Vila}},
  \bibinfo {author} {\bibfnamefont {C.-H.}\ \bibnamefont {Hsu}}, \bibinfo
  {author} {\bibfnamefont {X.}~\bibnamefont {Waintal}}, \bibinfo {author}
  {\bibfnamefont {V.~M.}\ \bibnamefont {Pereira}},\ and\ \bibinfo {author}
  {\bibfnamefont {S.}~\bibnamefont {Roche}},\ }\bibfield  {title} {\bibinfo
  {title} {Canted persistent spin texture and quantum spin hall effect in
  {WT}e$_{2}$},\ }\href {https://doi.org/10.1103/PhysRevLett.125.256603}
  {\bibfield  {journal} {\bibinfo  {journal} {Phys. Rev. Lett.}\ }\textbf
  {\bibinfo {volume} {125}},\ \bibinfo {pages} {256603} (\bibinfo {year}
  {2020})}\BibitemShut {NoStop}%
\bibitem [{\citenamefont {Tan}\ \emph {et~al.}(2021)\citenamefont {Tan},
  \citenamefont {Deng}, \citenamefont {Zheng}, \citenamefont {Xiang},
  \citenamefont {Albarakati}, \citenamefont {Algarni}, \citenamefont {Farrar},
  \citenamefont {Alzahrani}, \citenamefont {Partridge}, \citenamefont {Yi},
  \citenamefont {Hamilton}, \citenamefont {Wang},\ and\ \citenamefont
  {Wang}}]{SC_NL2021}%
  \BibitemOpen
  \bibfield  {author} {\bibinfo {author} {\bibfnamefont {C.}~\bibnamefont
  {Tan}}, \bibinfo {author} {\bibfnamefont {M.-X.}\ \bibnamefont {Deng}},
  \bibinfo {author} {\bibfnamefont {G.}~\bibnamefont {Zheng}}, \bibinfo
  {author} {\bibfnamefont {F.}~\bibnamefont {Xiang}}, \bibinfo {author}
  {\bibfnamefont {S.}~\bibnamefont {Albarakati}}, \bibinfo {author}
  {\bibfnamefont {M.}~\bibnamefont {Algarni}}, \bibinfo {author} {\bibfnamefont
  {L.}~\bibnamefont {Farrar}}, \bibinfo {author} {\bibfnamefont
  {S.}~\bibnamefont {Alzahrani}}, \bibinfo {author} {\bibfnamefont
  {J.}~\bibnamefont {Partridge}}, \bibinfo {author} {\bibfnamefont {J.~B.}\
  \bibnamefont {Yi}}, \bibinfo {author} {\bibfnamefont {A.~R.}\ \bibnamefont
  {Hamilton}}, \bibinfo {author} {\bibfnamefont {R.-Q.}\ \bibnamefont {Wang}},\
  and\ \bibinfo {author} {\bibfnamefont {L.}~\bibnamefont {Wang}},\ }\bibfield
  {title} {\bibinfo {title} {Spin-momentum locking induced anisotropic
  magnetoresistance in monolayer {WT}e$_{2}$},\ }\href
  {https://doi.org/10.1021/acs.nanolett.1c02329} {\bibfield  {journal}
  {\bibinfo  {journal} {Nano Lett.}\ }\textbf {\bibinfo {volume} {21}},\
  \bibinfo {pages} {9005–9011} (\bibinfo {year} {2021})}\BibitemShut
  {NoStop}%
\bibitem [{\citenamefont {Guo}\ \emph {et~al.}(2022)\citenamefont {Guo},
  \citenamefont {Hu}, \citenamefont {Putzke}, \citenamefont {Diaz},
  \citenamefont {Huang}, \citenamefont {Manna}, \citenamefont {Fan},
  \citenamefont {Shekhar}, \citenamefont {Sun}, \citenamefont {Felser},
  \citenamefont {Liu}, \citenamefont {Bernevig},\ and\ \citenamefont
  {Moll}}]{QS_NP2022}%
  \BibitemOpen
  \bibfield  {author} {\bibinfo {author} {\bibfnamefont {C.}~\bibnamefont
  {Guo}}, \bibinfo {author} {\bibfnamefont {L.}~\bibnamefont {Hu}}, \bibinfo
  {author} {\bibfnamefont {C.}~\bibnamefont {Putzke}}, \bibinfo {author}
  {\bibfnamefont {J.}~\bibnamefont {Diaz}}, \bibinfo {author} {\bibfnamefont
  {X.}~\bibnamefont {Huang}}, \bibinfo {author} {\bibfnamefont
  {K.}~\bibnamefont {Manna}}, \bibinfo {author} {\bibfnamefont {F.-R.}\
  \bibnamefont {Fan}}, \bibinfo {author} {\bibfnamefont {C.}~\bibnamefont
  {Shekhar}}, \bibinfo {author} {\bibfnamefont {Y.}~\bibnamefont {Sun}},
  \bibinfo {author} {\bibfnamefont {C.}~\bibnamefont {Felser}}, \bibinfo
  {author} {\bibfnamefont {C.}~\bibnamefont {Liu}}, \bibinfo {author}
  {\bibfnamefont {B.~A.}\ \bibnamefont {Bernevig}},\ and\ \bibinfo {author}
  {\bibfnamefont {P.~J.~W.}\ \bibnamefont {Moll}},\ }\bibfield  {title}
  {\bibinfo {title} {Quasi-symmetry-protected topology in a semi-metal},\
  }\href {https://doi.org/10.1038/s41567-022-01604-0} {\bibfield  {journal}
  {\bibinfo  {journal} {Nat. Phys.}\ }\textbf {\bibinfo {volume} {18}},\
  \bibinfo {pages} {813–818} (\bibinfo {year} {2022})}\BibitemShut {NoStop}%
\bibitem [{\citenamefont {Li}\ \emph {et~al.}(2024)\citenamefont {Li},
  \citenamefont {Zhang}, \citenamefont {Liu},\ and\ \citenamefont
  {Liu}}]{QS_PRL2024}%
  \BibitemOpen
  \bibfield  {author} {\bibinfo {author} {\bibfnamefont {J.}~\bibnamefont
  {Li}}, \bibinfo {author} {\bibfnamefont {A.}~\bibnamefont {Zhang}}, \bibinfo
  {author} {\bibfnamefont {Y.}~\bibnamefont {Liu}},\ and\ \bibinfo {author}
  {\bibfnamefont {Q.}~\bibnamefont {Liu}},\ }\bibfield  {title} {\bibinfo
  {title} {Group theory on quasisymmetry and protected near degeneracy},\
  }\href {https://doi.org/10.1103/PhysRevLett.133.026402} {\bibfield  {journal}
  {\bibinfo  {journal} {Phys. Rev. Lett.}\ }\textbf {\bibinfo {volume} {133}},\
  \bibinfo {pages} {026402} (\bibinfo {year} {2024})}\BibitemShut {NoStop}%
\bibitem [{\citenamefont {Liu}\ \emph {et~al.}(2024)\citenamefont {Liu},
  \citenamefont {Liu}, \citenamefont {Li}, \citenamefont {Wu},\ and\
  \citenamefont {Liu}}]{QS_PRB2024}%
  \BibitemOpen
  \bibfield  {author} {\bibinfo {author} {\bibfnamefont {L.}~\bibnamefont
  {Liu}}, \bibinfo {author} {\bibfnamefont {Y.}~\bibnamefont {Liu}}, \bibinfo
  {author} {\bibfnamefont {J.}~\bibnamefont {Li}}, \bibinfo {author}
  {\bibfnamefont {H.}~\bibnamefont {Wu}},\ and\ \bibinfo {author}
  {\bibfnamefont {Q.}~\bibnamefont {Liu}},\ }\bibfield  {title} {\bibinfo
  {title} {Quantum spin hall effect protected by spin {U(1)} quasisymmetry},\
  }\href {https://doi.org/10.1103/PhysRevB.110.L161104} {\bibfield  {journal}
  {\bibinfo  {journal} {Phys. Rev. B}\ }\textbf {\bibinfo {volume} {110}},\
  \bibinfo {pages} {L161104} (\bibinfo {year} {2024})}\BibitemShut {NoStop}%
\bibitem [{\citenamefont {Verma}\ \emph {et~al.}(2024)\citenamefont {Verma},
  \citenamefont {Huang},\ and\ \citenamefont {Singh}}]{QS_SbVerma2024}%
  \BibitemOpen
  \bibfield  {author} {\bibinfo {author} {\bibfnamefont {R.}~\bibnamefont
  {Verma}}, \bibinfo {author} {\bibfnamefont {S.-M.}\ \bibnamefont {Huang}},\
  and\ \bibinfo {author} {\bibfnamefont {B.}~\bibnamefont {Singh}},\ }\bibfield
   {title} {\bibinfo {title} {Atomically thin obstructed atomic insulators with
  robust edge modes and quantized spin hall effect},\ }\href
  {https://doi.org/10.1103/PhysRevB.110.165122} {\bibfield  {journal} {\bibinfo
   {journal} {Phys. Rev. B}\ }\textbf {\bibinfo {volume} {110}},\ \bibinfo
  {pages} {165122} (\bibinfo {year} {2024})}\BibitemShut {NoStop}%
\bibitem [{\citenamefont {Du}\ \emph {et~al.}(2015)\citenamefont {Du},
  \citenamefont {Knez}, \citenamefont {Sullivan},\ and\ \citenamefont
  {Du}}]{QSH_InAs2015}%
  \BibitemOpen
  \bibfield  {author} {\bibinfo {author} {\bibfnamefont {L.}~\bibnamefont
  {Du}}, \bibinfo {author} {\bibfnamefont {I.}~\bibnamefont {Knez}}, \bibinfo
  {author} {\bibfnamefont {G.}~\bibnamefont {Sullivan}},\ and\ \bibinfo
  {author} {\bibfnamefont {R.-R.}\ \bibnamefont {Du}},\ }\bibfield  {title}
  {\bibinfo {title} {Robust helical edge transport in gated
  $\mathrm{InAs}/\mathrm{GaSb}$ bilayers},\ }\href
  {https://doi.org/10.1103/PhysRevLett.114.096802} {\bibfield  {journal}
  {\bibinfo  {journal} {Phys. Rev. Lett.}\ }\textbf {\bibinfo {volume} {114}},\
  \bibinfo {pages} {096802} (\bibinfo {year} {2015})}\BibitemShut {NoStop}%
\bibitem [{\citenamefont {Wu}\ \emph {et~al.}(2018{\natexlab{a}})\citenamefont
  {Wu}, \citenamefont {Fatemi}, \citenamefont {Gibson}, \citenamefont
  {Watanabe}, \citenamefont {Taniguchi}, \citenamefont {Cava},\ and\
  \citenamefont {Jarillo-Herrero}}]{QSH_T100K2018}%
  \BibitemOpen
  \bibfield  {author} {\bibinfo {author} {\bibfnamefont {S.}~\bibnamefont
  {Wu}}, \bibinfo {author} {\bibfnamefont {V.}~\bibnamefont {Fatemi}}, \bibinfo
  {author} {\bibfnamefont {Q.~D.}\ \bibnamefont {Gibson}}, \bibinfo {author}
  {\bibfnamefont {K.}~\bibnamefont {Watanabe}}, \bibinfo {author}
  {\bibfnamefont {T.}~\bibnamefont {Taniguchi}}, \bibinfo {author}
  {\bibfnamefont {R.~J.}\ \bibnamefont {Cava}},\ and\ \bibinfo {author}
  {\bibfnamefont {P.}~\bibnamefont {Jarillo-Herrero}},\ }\bibfield  {title}
  {\bibinfo {title} {Observation of the quantum spin hall effect up to 100
  kelvin in a monolayer crystal},\ }\href
  {https://doi.org/10.1126/science.aan6003} {\bibfield  {journal} {\bibinfo
  {journal} {Science}\ }\textbf {\bibinfo {volume} {359}},\ \bibinfo {pages}
  {76–79} (\bibinfo {year} {2018}{\natexlab{a}})}\BibitemShut {NoStop}%
\bibitem [{\citenamefont {Zhao}\ \emph {et~al.}(2024)\citenamefont {Zhao},
  \citenamefont {Kang}, \citenamefont {Zhang}, \citenamefont {Kn\"{u}ppel},
  \citenamefont {Tao}, \citenamefont {Li}, \citenamefont {Tschirhart},
  \citenamefont {Redekop}, \citenamefont {Watanabe}, \citenamefont {Taniguchi},
  \citenamefont {Young}, \citenamefont {Shan},\ and\ \citenamefont
  {Mak}}]{QSH_Moire2024}%
  \BibitemOpen
  \bibfield  {author} {\bibinfo {author} {\bibfnamefont {W.}~\bibnamefont
  {Zhao}}, \bibinfo {author} {\bibfnamefont {K.}~\bibnamefont {Kang}}, \bibinfo
  {author} {\bibfnamefont {Y.}~\bibnamefont {Zhang}}, \bibinfo {author}
  {\bibfnamefont {P.}~\bibnamefont {Kn\"{u}ppel}}, \bibinfo {author}
  {\bibfnamefont {Z.}~\bibnamefont {Tao}}, \bibinfo {author} {\bibfnamefont
  {L.}~\bibnamefont {Li}}, \bibinfo {author} {\bibfnamefont {C.~L.}\
  \bibnamefont {Tschirhart}}, \bibinfo {author} {\bibfnamefont
  {E.}~\bibnamefont {Redekop}}, \bibinfo {author} {\bibfnamefont
  {K.}~\bibnamefont {Watanabe}}, \bibinfo {author} {\bibfnamefont
  {T.}~\bibnamefont {Taniguchi}}, \bibinfo {author} {\bibfnamefont {A.~F.}\
  \bibnamefont {Young}}, \bibinfo {author} {\bibfnamefont {J.}~\bibnamefont
  {Shan}},\ and\ \bibinfo {author} {\bibfnamefont {K.~F.}\ \bibnamefont
  {Mak}},\ }\bibfield  {title} {\bibinfo {title} {Realization of the haldane
  chern insulator in a moiré lattice},\ }\href
  {https://doi.org/10.1038/s41567-023-02284-0} {\bibfield  {journal} {\bibinfo
  {journal} {Nat. Phys.}\ }\textbf {\bibinfo {volume} {20}},\ \bibinfo {pages}
  {275–280} (\bibinfo {year} {2024})}\BibitemShut {NoStop}%
\bibitem [{\citenamefont {Kang}\ \emph {et~al.}(2024)\citenamefont {Kang},
  \citenamefont {Qiu}, \citenamefont {Watanabe}, \citenamefont {Taniguchi},
  \citenamefont {Shan},\ and\ \citenamefont {Mak}}]{QSH_moireWSe22024}%
  \BibitemOpen
  \bibfield  {author} {\bibinfo {author} {\bibfnamefont {K.}~\bibnamefont
  {Kang}}, \bibinfo {author} {\bibfnamefont {Y.}~\bibnamefont {Qiu}}, \bibinfo
  {author} {\bibfnamefont {K.}~\bibnamefont {Watanabe}}, \bibinfo {author}
  {\bibfnamefont {T.}~\bibnamefont {Taniguchi}}, \bibinfo {author}
  {\bibfnamefont {J.}~\bibnamefont {Shan}},\ and\ \bibinfo {author}
  {\bibfnamefont {K.~F.}\ \bibnamefont {Mak}},\ }\bibfield  {title} {\bibinfo
  {title} {Double quantum spin hall phase in moiré {WSe$_2$}},\ }\href
  {https://doi.org/10.1021/acs.nanolett.4c05308} {\bibfield  {journal}
  {\bibinfo  {journal} {Nano Lett.}\ }\textbf {\bibinfo {volume} {24}},\
  \bibinfo {pages} {14901–14907} (\bibinfo {year} {2024})}\BibitemShut
  {NoStop}%
\bibitem [{\citenamefont {Tang}\ \emph {et~al.}(2024)\citenamefont {Tang},
  \citenamefont {Ding}, \citenamefont {Chen}, \citenamefont {Gao},
  \citenamefont {Qian}, \citenamefont {Huang}, \citenamefont {Sun},
  \citenamefont {Han}, \citenamefont {Strasser}, \citenamefont {Li},
  \citenamefont {Geiwitz}, \citenamefont {Shehabeldin}, \citenamefont
  {Belosevich}, \citenamefont {Wang}, \citenamefont {Wang}, \citenamefont
  {Watanabe}, \citenamefont {Taniguchi}, \citenamefont {Bell}, \citenamefont
  {Wang}, \citenamefont {Fu}, \citenamefont {Zhang}, \citenamefont {Qian},
  \citenamefont {Burch}, \citenamefont {Shi}, \citenamefont {Ni}, \citenamefont
  {Chang}, \citenamefont {Xu},\ and\ \citenamefont {Ma}}]{DQSH_TaIrTe42024}%
  \BibitemOpen
  \bibfield  {author} {\bibinfo {author} {\bibfnamefont {J.}~\bibnamefont
  {Tang}}, \bibinfo {author} {\bibfnamefont {T.~S.}\ \bibnamefont {Ding}},
  \bibinfo {author} {\bibfnamefont {H.}~\bibnamefont {Chen}}, \bibinfo {author}
  {\bibfnamefont {A.}~\bibnamefont {Gao}}, \bibinfo {author} {\bibfnamefont
  {T.}~\bibnamefont {Qian}}, \bibinfo {author} {\bibfnamefont {Z.}~\bibnamefont
  {Huang}}, \bibinfo {author} {\bibfnamefont {Z.}~\bibnamefont {Sun}}, \bibinfo
  {author} {\bibfnamefont {X.}~\bibnamefont {Han}}, \bibinfo {author}
  {\bibfnamefont {A.}~\bibnamefont {Strasser}}, \bibinfo {author}
  {\bibfnamefont {J.}~\bibnamefont {Li}}, \bibinfo {author} {\bibfnamefont
  {M.}~\bibnamefont {Geiwitz}}, \bibinfo {author} {\bibfnamefont
  {M.}~\bibnamefont {Shehabeldin}}, \bibinfo {author} {\bibfnamefont
  {V.}~\bibnamefont {Belosevich}}, \bibinfo {author} {\bibfnamefont
  {Z.}~\bibnamefont {Wang}}, \bibinfo {author} {\bibfnamefont {Y.}~\bibnamefont
  {Wang}}, \bibinfo {author} {\bibfnamefont {K.}~\bibnamefont {Watanabe}},
  \bibinfo {author} {\bibfnamefont {T.}~\bibnamefont {Taniguchi}}, \bibinfo
  {author} {\bibfnamefont {D.~C.}\ \bibnamefont {Bell}}, \bibinfo {author}
  {\bibfnamefont {Z.}~\bibnamefont {Wang}}, \bibinfo {author} {\bibfnamefont
  {L.}~\bibnamefont {Fu}}, \bibinfo {author} {\bibfnamefont {Y.}~\bibnamefont
  {Zhang}}, \bibinfo {author} {\bibfnamefont {X.}~\bibnamefont {Qian}},
  \bibinfo {author} {\bibfnamefont {K.~S.}\ \bibnamefont {Burch}}, \bibinfo
  {author} {\bibfnamefont {Y.}~\bibnamefont {Shi}}, \bibinfo {author}
  {\bibfnamefont {N.}~\bibnamefont {Ni}}, \bibinfo {author} {\bibfnamefont
  {G.}~\bibnamefont {Chang}}, \bibinfo {author} {\bibfnamefont {S.-Y.}\
  \bibnamefont {Xu}},\ and\ \bibinfo {author} {\bibfnamefont {Q.}~\bibnamefont
  {Ma}},\ }\bibfield  {title} {\bibinfo {title} {Dual quantum spin hall
  insulator by density-tuned correlations in {TaIrTe$_4$}},\ }\href
  {https://doi.org/10.1038/s41586-024-07211-8} {\bibfield  {journal} {\bibinfo
  {journal} {Nature}\ }\textbf {\bibinfo {volume} {628}},\ \bibinfo {pages}
  {515–521} (\bibinfo {year} {2024})}\BibitemShut {NoStop}%
\bibitem [{\citenamefont {Tokura}(2022)}]{Tokura2022}%
  \BibitemOpen
  \bibfield  {author} {\bibinfo {author} {\bibfnamefont {Y.}~\bibnamefont
  {Tokura}},\ }\bibfield  {title} {\bibinfo {title} {Quantum materials at the
  crossroads of strong correlation and topology},\ }\href
  {https://doi.org/10.1038/s41563-022-01339-6} {\bibfield  {journal} {\bibinfo
  {journal} {Nat. Mater.}\ }\textbf {\bibinfo {volume} {21}},\ \bibinfo {pages}
  {971–973} (\bibinfo {year} {2022})}\BibitemShut {NoStop}%
\bibitem [{\citenamefont {Cai}\ \emph {et~al.}(2023)\citenamefont {Cai},
  \citenamefont {Anderson}, \citenamefont {Wang}, \citenamefont {Zhang},
  \citenamefont {Liu}, \citenamefont {Holtzmann}, \citenamefont {Zhang},
  \citenamefont {Fan}, \citenamefont {Taniguchi}, \citenamefont {Watanabe},
  \citenamefont {Ran}, \citenamefont {Cao}, \citenamefont {Fu}, \citenamefont
  {Xiao}, \citenamefont {Yao},\ and\ \citenamefont {Xu}}]{Cai2023}%
  \BibitemOpen
  \bibfield  {author} {\bibinfo {author} {\bibfnamefont {J.}~\bibnamefont
  {Cai}}, \bibinfo {author} {\bibfnamefont {E.}~\bibnamefont {Anderson}},
  \bibinfo {author} {\bibfnamefont {C.}~\bibnamefont {Wang}}, \bibinfo {author}
  {\bibfnamefont {X.}~\bibnamefont {Zhang}}, \bibinfo {author} {\bibfnamefont
  {X.}~\bibnamefont {Liu}}, \bibinfo {author} {\bibfnamefont {W.}~\bibnamefont
  {Holtzmann}}, \bibinfo {author} {\bibfnamefont {Y.}~\bibnamefont {Zhang}},
  \bibinfo {author} {\bibfnamefont {F.}~\bibnamefont {Fan}}, \bibinfo {author}
  {\bibfnamefont {T.}~\bibnamefont {Taniguchi}}, \bibinfo {author}
  {\bibfnamefont {K.}~\bibnamefont {Watanabe}}, \bibinfo {author}
  {\bibfnamefont {Y.}~\bibnamefont {Ran}}, \bibinfo {author} {\bibfnamefont
  {T.}~\bibnamefont {Cao}}, \bibinfo {author} {\bibfnamefont {L.}~\bibnamefont
  {Fu}}, \bibinfo {author} {\bibfnamefont {D.}~\bibnamefont {Xiao}}, \bibinfo
  {author} {\bibfnamefont {W.}~\bibnamefont {Yao}},\ and\ \bibinfo {author}
  {\bibfnamefont {X.}~\bibnamefont {Xu}},\ }\bibfield  {title} {\bibinfo
  {title} {Signatures of fractional quantum anomalous hall states in twisted
  {MoTe$_2$}},\ }\href {https://doi.org/10.1038/s41586-023-06289-w} {\bibfield
  {journal} {\bibinfo  {journal} {Nature}\ }\textbf {\bibinfo {volume} {622}},\
  \bibinfo {pages} {63–68} (\bibinfo {year} {2023})}\BibitemShut {NoStop}%
\bibitem [{\citenamefont {Dzero}\ \emph {et~al.}(2010)\citenamefont {Dzero},
  \citenamefont {Sun}, \citenamefont {Galitski},\ and\ \citenamefont
  {Coleman}}]{Kondo_PRL2010}%
  \BibitemOpen
  \bibfield  {author} {\bibinfo {author} {\bibfnamefont {M.}~\bibnamefont
  {Dzero}}, \bibinfo {author} {\bibfnamefont {K.}~\bibnamefont {Sun}}, \bibinfo
  {author} {\bibfnamefont {V.}~\bibnamefont {Galitski}},\ and\ \bibinfo
  {author} {\bibfnamefont {P.}~\bibnamefont {Coleman}},\ }\bibfield  {title}
  {\bibinfo {title} {Topological kondo insulators},\ }\href
  {https://doi.org/10.1103/PhysRevLett.104.106408} {\bibfield  {journal}
  {\bibinfo  {journal} {Phys. Rev. Lett.}\ }\textbf {\bibinfo {volume} {104}},\
  \bibinfo {pages} {106408} (\bibinfo {year} {2010})}\BibitemShut {NoStop}%
\bibitem [{\citenamefont {Zhang}\ \emph {et~al.}(2022)\citenamefont {Zhang},
  \citenamefont {Singh}, \citenamefont {Lane}, \citenamefont {Kidd},
  \citenamefont {Zhang}, \citenamefont {Barbiellini}, \citenamefont
  {Markiewicz}, \citenamefont {Bansil},\ and\ \citenamefont
  {Sun}}]{SmB6_Singh2022}%
  \BibitemOpen
  \bibfield  {author} {\bibinfo {author} {\bibfnamefont {R.}~\bibnamefont
  {Zhang}}, \bibinfo {author} {\bibfnamefont {B.}~\bibnamefont {Singh}},
  \bibinfo {author} {\bibfnamefont {C.}~\bibnamefont {Lane}}, \bibinfo {author}
  {\bibfnamefont {J.}~\bibnamefont {Kidd}}, \bibinfo {author} {\bibfnamefont
  {Y.}~\bibnamefont {Zhang}}, \bibinfo {author} {\bibfnamefont
  {B.}~\bibnamefont {Barbiellini}}, \bibinfo {author} {\bibfnamefont {R.~S.}\
  \bibnamefont {Markiewicz}}, \bibinfo {author} {\bibfnamefont
  {A.}~\bibnamefont {Bansil}},\ and\ \bibinfo {author} {\bibfnamefont
  {J.}~\bibnamefont {Sun}},\ }\bibfield  {title} {\bibinfo {title} {Critical
  role of magnetic moments in heavy-fermion materials: Revisiting
  {SmB}$_{6}$},\ }\href {https://doi.org/10.1103/PhysRevB.105.195134}
  {\bibfield  {journal} {\bibinfo  {journal} {Phys. Rev. B}\ }\textbf {\bibinfo
  {volume} {105}},\ \bibinfo {pages} {195134} (\bibinfo {year}
  {2022})}\BibitemShut {NoStop}%
\bibitem [{\citenamefont {Chang}\ \emph {et~al.}(2017)\citenamefont {Chang},
  \citenamefont {Xu}, \citenamefont {Zhou}, \citenamefont {Huang},
  \citenamefont {Singh}, \citenamefont {Wang}, \citenamefont {Belopolski},
  \citenamefont {Yin}, \citenamefont {Zhang}, \citenamefont {Bansil},
  \citenamefont {Lin},\ and\ \citenamefont {Hasan}}]{MWSM_BSPRL2017}%
  \BibitemOpen
  \bibfield  {author} {\bibinfo {author} {\bibfnamefont {G.}~\bibnamefont
  {Chang}}, \bibinfo {author} {\bibfnamefont {S.-Y.}\ \bibnamefont {Xu}},
  \bibinfo {author} {\bibfnamefont {X.}~\bibnamefont {Zhou}}, \bibinfo {author}
  {\bibfnamefont {S.-M.}\ \bibnamefont {Huang}}, \bibinfo {author}
  {\bibfnamefont {B.}~\bibnamefont {Singh}}, \bibinfo {author} {\bibfnamefont
  {B.}~\bibnamefont {Wang}}, \bibinfo {author} {\bibfnamefont {I.}~\bibnamefont
  {Belopolski}}, \bibinfo {author} {\bibfnamefont {J.}~\bibnamefont {Yin}},
  \bibinfo {author} {\bibfnamefont {S.}~\bibnamefont {Zhang}}, \bibinfo
  {author} {\bibfnamefont {A.}~\bibnamefont {Bansil}}, \bibinfo {author}
  {\bibfnamefont {H.}~\bibnamefont {Lin}},\ and\ \bibinfo {author}
  {\bibfnamefont {M.~Z.}\ \bibnamefont {Hasan}},\ }\bibfield  {title} {\bibinfo
  {title} {Topological hopf and chain link semimetal states and their
  application to ${\mathrm{co}}_{2}\mathrm{Mn}\text{G}\text{a}$},\ }\href
  {https://doi.org/10.1103/PhysRevLett.119.156401} {\bibfield  {journal}
  {\bibinfo  {journal} {Phys. Rev. Lett.}\ }\textbf {\bibinfo {volume} {119}},\
  \bibinfo {pages} {156401} (\bibinfo {year} {2017})}\BibitemShut {NoStop}%
\bibitem [{\citenamefont {Mardanya}\ \emph {et~al.}(2024)\citenamefont
  {Mardanya}, \citenamefont {Kargarian}, \citenamefont {Verma}, \citenamefont
  {Chang}, \citenamefont {Chowdhury}, \citenamefont {Lin}, \citenamefont
  {Bansil}, \citenamefont {Agarwal},\ and\ \citenamefont
  {Singh}}]{CP_RPMSingh2024}%
  \BibitemOpen
  \bibfield  {author} {\bibinfo {author} {\bibfnamefont {S.}~\bibnamefont
  {Mardanya}}, \bibinfo {author} {\bibfnamefont {M.}~\bibnamefont {Kargarian}},
  \bibinfo {author} {\bibfnamefont {R.}~\bibnamefont {Verma}}, \bibinfo
  {author} {\bibfnamefont {T.-R.}\ \bibnamefont {Chang}}, \bibinfo {author}
  {\bibfnamefont {S.}~\bibnamefont {Chowdhury}}, \bibinfo {author}
  {\bibfnamefont {H.}~\bibnamefont {Lin}}, \bibinfo {author} {\bibfnamefont
  {A.}~\bibnamefont {Bansil}}, \bibinfo {author} {\bibfnamefont
  {A.}~\bibnamefont {Agarwal}},\ and\ \bibinfo {author} {\bibfnamefont
  {B.}~\bibnamefont {Singh}},\ }\bibfield  {title} {\bibinfo {title}
  {Unconventional superconducting pairing in a b20 multifold weyl fermion
  semimetal},\ }\href {https://doi.org/10.1103/PhysRevMaterials.8.L091801}
  {\bibfield  {journal} {\bibinfo  {journal} {Phys. Rev. Mater.}\ }\textbf
  {\bibinfo {volume} {8}},\ \bibinfo {pages} {L091801} (\bibinfo {year}
  {2024})}\BibitemShut {NoStop}%
\bibitem [{\citenamefont {Markiewicz}\ \emph {et~al.}(2023)\citenamefont
  {Markiewicz}, \citenamefont {Singh}, \citenamefont {Lane},\ and\
  \citenamefont {Bansil}}]{CP_CPSingh2023}%
  \BibitemOpen
  \bibfield  {author} {\bibinfo {author} {\bibfnamefont {R.~S.}\ \bibnamefont
  {Markiewicz}}, \bibinfo {author} {\bibfnamefont {B.}~\bibnamefont {Singh}},
  \bibinfo {author} {\bibfnamefont {C.}~\bibnamefont {Lane}},\ and\ \bibinfo
  {author} {\bibfnamefont {A.}~\bibnamefont {Bansil}},\ }\bibfield  {title}
  {\bibinfo {title} {Investigating the {Cuprates} as a platform for high-order
  {Van} {Hove} singularities and flat-band physics},\ }\href
  {https://doi.org/10.1038/s42005-023-01373-z} {\bibfield  {journal} {\bibinfo
  {journal} {Commun. Phys.}\ }\textbf {\bibinfo {volume} {6}},\ \bibinfo
  {pages} {292} (\bibinfo {year} {2023})}\BibitemShut {NoStop}%
\bibitem [{\citenamefont {Hu}\ \emph {et~al.}(2020)\citenamefont {Hu},
  \citenamefont {Ding}, \citenamefont {Gordon}, \citenamefont {Ghosh},
  \citenamefont {Tien}, \citenamefont {Li}, \citenamefont {Linn}, \citenamefont
  {Lien}, \citenamefont {Huang}, \citenamefont {Mackey}, \citenamefont {Liu},
  \citenamefont {Reddy}, \citenamefont {Singh}, \citenamefont {Agarwal},
  \citenamefont {Bansil}, \citenamefont {Song}, \citenamefont {Li},
  \citenamefont {Xu}, \citenamefont {Lin}, \citenamefont {Cao}, \citenamefont
  {Chang}, \citenamefont {Dessau},\ and\ \citenamefont {Ni}}]{Hu2020}%
  \BibitemOpen
  \bibfield  {author} {\bibinfo {author} {\bibfnamefont {C.}~\bibnamefont
  {Hu}}, \bibinfo {author} {\bibfnamefont {L.}~\bibnamefont {Ding}}, \bibinfo
  {author} {\bibfnamefont {K.~N.}\ \bibnamefont {Gordon}}, \bibinfo {author}
  {\bibfnamefont {B.}~\bibnamefont {Ghosh}}, \bibinfo {author} {\bibfnamefont
  {H.-J.}\ \bibnamefont {Tien}}, \bibinfo {author} {\bibfnamefont
  {H.}~\bibnamefont {Li}}, \bibinfo {author} {\bibfnamefont {A.~G.}\
  \bibnamefont {Linn}}, \bibinfo {author} {\bibfnamefont {S.-W.}\ \bibnamefont
  {Lien}}, \bibinfo {author} {\bibfnamefont {C.-Y.}\ \bibnamefont {Huang}},
  \bibinfo {author} {\bibfnamefont {S.}~\bibnamefont {Mackey}}, \bibinfo
  {author} {\bibfnamefont {J.}~\bibnamefont {Liu}}, \bibinfo {author}
  {\bibfnamefont {P.~V.~S.}\ \bibnamefont {Reddy}}, \bibinfo {author}
  {\bibfnamefont {B.}~\bibnamefont {Singh}}, \bibinfo {author} {\bibfnamefont
  {A.}~\bibnamefont {Agarwal}}, \bibinfo {author} {\bibfnamefont
  {A.}~\bibnamefont {Bansil}}, \bibinfo {author} {\bibfnamefont
  {M.}~\bibnamefont {Song}}, \bibinfo {author} {\bibfnamefont {D.}~\bibnamefont
  {Li}}, \bibinfo {author} {\bibfnamefont {S.-Y.}\ \bibnamefont {Xu}}, \bibinfo
  {author} {\bibfnamefont {H.}~\bibnamefont {Lin}}, \bibinfo {author}
  {\bibfnamefont {H.}~\bibnamefont {Cao}}, \bibinfo {author} {\bibfnamefont
  {T.-R.}\ \bibnamefont {Chang}}, \bibinfo {author} {\bibfnamefont
  {D.}~\bibnamefont {Dessau}},\ and\ \bibinfo {author} {\bibfnamefont
  {N.}~\bibnamefont {Ni}},\ }\bibfield  {title} {\bibinfo {title} {Realization
  of an intrinsic ferromagnetic topological state in {MnBi$_8$Te$_{13}$}},\
  }\href {https://doi.org/10.1126/sciadv.aba4275} {\bibfield  {journal}
  {\bibinfo  {journal} {Sci. Adv.}\ }\textbf {\bibinfo {volume} {6}},\ \bibinfo
  {pages} {eaba4275} (\bibinfo {year} {2020})}\BibitemShut {NoStop}%
\bibitem [{\citenamefont {Patra}\ \emph {et~al.}(2025)\citenamefont {Patra},
  \citenamefont {Mukherjee},\ and\ \citenamefont {Singh}}]{Kagome_Patra2025}%
  \BibitemOpen
  \bibfield  {author} {\bibinfo {author} {\bibfnamefont {B.}~\bibnamefont
  {Patra}}, \bibinfo {author} {\bibfnamefont {A.}~\bibnamefont {Mukherjee}},\
  and\ \bibinfo {author} {\bibfnamefont {B.}~\bibnamefont {Singh}},\ }\bibfield
   {title} {\bibinfo {title} {High-order van hove singularities and nematic
  instability in the kagome superconductor {CsTi}$_{3}${Bi}$_{5}$},\ }\href
  {https://doi.org/10.1103/PhysRevB.111.045135} {\bibfield  {journal} {\bibinfo
   {journal} {Phys. Rev. B}\ }\textbf {\bibinfo {volume} {111}},\ \bibinfo
  {pages} {045135} (\bibinfo {year} {2025})}\BibitemShut {NoStop}%
\bibitem [{\citenamefont {Hong}\ \emph {et~al.}(2020)\citenamefont {Hong},
  \citenamefont {Liu}, \citenamefont {Wang}, \citenamefont {Zhou},
  \citenamefont {Ma}, \citenamefont {Xu}, \citenamefont {Feng}, \citenamefont
  {Chen}, \citenamefont {Chen}, \citenamefont {Sun}, \citenamefont {Chen},
  \citenamefont {Cheng},\ and\ \citenamefont {Ren}}]{Hong2020}%
  \BibitemOpen
  \bibfield  {author} {\bibinfo {author} {\bibfnamefont {Y.-L.}\ \bibnamefont
  {Hong}}, \bibinfo {author} {\bibfnamefont {Z.}~\bibnamefont {Liu}}, \bibinfo
  {author} {\bibfnamefont {L.}~\bibnamefont {Wang}}, \bibinfo {author}
  {\bibfnamefont {T.}~\bibnamefont {Zhou}}, \bibinfo {author} {\bibfnamefont
  {W.}~\bibnamefont {Ma}}, \bibinfo {author} {\bibfnamefont {C.}~\bibnamefont
  {Xu}}, \bibinfo {author} {\bibfnamefont {S.}~\bibnamefont {Feng}}, \bibinfo
  {author} {\bibfnamefont {L.}~\bibnamefont {Chen}}, \bibinfo {author}
  {\bibfnamefont {M.-L.}\ \bibnamefont {Chen}}, \bibinfo {author}
  {\bibfnamefont {D.-M.}\ \bibnamefont {Sun}}, \bibinfo {author} {\bibfnamefont
  {X.-Q.}\ \bibnamefont {Chen}}, \bibinfo {author} {\bibfnamefont {H.-M.}\
  \bibnamefont {Cheng}},\ and\ \bibinfo {author} {\bibfnamefont
  {W.}~\bibnamefont {Ren}},\ }\bibfield  {title} {\bibinfo {title} {Chemical
  vapor deposition of layered two-dimensional {MoSi$_2$N$_4$} materials},\
  }\href {https://doi.org/10.1126/science.abb7023} {\bibfield  {journal}
  {\bibinfo  {journal} {Science}\ }\textbf {\bibinfo {volume} {369}},\ \bibinfo
  {pages} {670–674} (\bibinfo {year} {2020})}\BibitemShut {NoStop}%
\bibitem [{\citenamefont {Latychevskaia}\ \emph {et~al.}(2024)\citenamefont
  {Latychevskaia}, \citenamefont {Bandurin},\ and\ \citenamefont
  {Novoselov}}]{Novesolov2024}%
  \BibitemOpen
  \bibfield  {author} {\bibinfo {author} {\bibfnamefont {T.}~\bibnamefont
  {Latychevskaia}}, \bibinfo {author} {\bibfnamefont {D.~A.}\ \bibnamefont
  {Bandurin}},\ and\ \bibinfo {author} {\bibfnamefont {K.~S.}\ \bibnamefont
  {Novoselov}},\ }\bibfield  {title} {\bibinfo {title} {A new family of
  septuple-layer {2D} materials of {MoSi$_2$N$_4$}-like crystals},\ }\href
  {https://doi.org/10.1038/s42254-024-00728-x} {\bibfield  {journal} {\bibinfo
  {journal} {Nat. Rev. Phys.}\ }\textbf {\bibinfo {volume} {6}},\ \bibinfo
  {pages} {426–438} (\bibinfo {year} {2024})}\BibitemShut {NoStop}%
\bibitem [{\citenamefont {Islam}\ \emph {et~al.}(2022)\citenamefont {Islam},
  \citenamefont {Verma}, \citenamefont {Ghosh}, \citenamefont {Muhammad},
  \citenamefont {Bansil}, \citenamefont {Autieri},\ and\ \citenamefont
  {Singh}}]{QSH_Verma2022}%
  \BibitemOpen
  \bibfield  {author} {\bibinfo {author} {\bibfnamefont {R.}~\bibnamefont
  {Islam}}, \bibinfo {author} {\bibfnamefont {R.}~\bibnamefont {Verma}},
  \bibinfo {author} {\bibfnamefont {B.}~\bibnamefont {Ghosh}}, \bibinfo
  {author} {\bibfnamefont {Z.}~\bibnamefont {Muhammad}}, \bibinfo {author}
  {\bibfnamefont {A.}~\bibnamefont {Bansil}}, \bibinfo {author} {\bibfnamefont
  {C.}~\bibnamefont {Autieri}},\ and\ \bibinfo {author} {\bibfnamefont
  {B.}~\bibnamefont {Singh}},\ }\bibfield  {title} {\bibinfo {title}
  {Switchable large-gap quantum spin hall state in the two-dimensional
  {MSi$_2$N$_4$} class of materials},\ }\href
  {https://doi.org/10.1103/PhysRevB.106.245149} {\bibfield  {journal} {\bibinfo
   {journal} {Phys. Rev. B}\ }\textbf {\bibinfo {volume} {106}},\ \bibinfo
  {pages} {245149} (\bibinfo {year} {2022})}\BibitemShut {NoStop}%
\bibitem [{\citenamefont {Islam}\ \emph {et~al.}(2023)\citenamefont {Islam},
  \citenamefont {Hussain}, \citenamefont {Verma}, \citenamefont
  {Talezadehlari}, \citenamefont {Muhammad}, \citenamefont {Singh},\ and\
  \citenamefont {Autieri}}]{Islam2023}%
  \BibitemOpen
  \bibfield  {author} {\bibinfo {author} {\bibfnamefont {R.}~\bibnamefont
  {Islam}}, \bibinfo {author} {\bibfnamefont {G.}~\bibnamefont {Hussain}},
  \bibinfo {author} {\bibfnamefont {R.}~\bibnamefont {Verma}}, \bibinfo
  {author} {\bibfnamefont {M.~S.}\ \bibnamefont {Talezadehlari}}, \bibinfo
  {author} {\bibfnamefont {Z.}~\bibnamefont {Muhammad}}, \bibinfo {author}
  {\bibfnamefont {B.}~\bibnamefont {Singh}},\ and\ \bibinfo {author}
  {\bibfnamefont {C.}~\bibnamefont {Autieri}},\ }\bibfield  {title} {\bibinfo
  {title} {Fast electrically switchable large gap quantum spin hall states in
  {MGe$_2$Z$_4$}},\ }\href {http://dx.doi.org/10.1002/aelm.202300156}
  {\bibfield  {journal} {\bibinfo  {journal} {Adv. Electron. Mater.}\ }\textbf
  {\bibinfo {volume} {9}} (\bibinfo {year} {2023})}\BibitemShut {NoStop}%
\bibitem [{\citenamefont {Hohenberg}\ and\ \citenamefont
  {Kohn}(1964)}]{Hohen1964}%
  \BibitemOpen
  \bibfield  {author} {\bibinfo {author} {\bibfnamefont {P.}~\bibnamefont
  {Hohenberg}}\ and\ \bibinfo {author} {\bibfnamefont {W.}~\bibnamefont
  {Kohn}},\ }\bibfield  {title} {\bibinfo {title} {Inhomogeneous electron
  gas},\ }\href {https://doi.org/10.1103/PhysRev.136.B864} {\bibfield
  {journal} {\bibinfo  {journal} {Phys. Rev.}\ }\textbf {\bibinfo {volume}
  {136}},\ \bibinfo {pages} {B864} (\bibinfo {year} {1964})}\BibitemShut
  {NoStop}%
\bibitem [{\citenamefont {Kresse}\ and\ \citenamefont
  {Furthm\"uller}(1996)}]{Kresse1996}%
  \BibitemOpen
  \bibfield  {author} {\bibinfo {author} {\bibfnamefont {G.}~\bibnamefont
  {Kresse}}\ and\ \bibinfo {author} {\bibfnamefont {J.}~\bibnamefont
  {Furthm\"uller}},\ }\bibfield  {title} {\bibinfo {title} {Efficient iterative
  schemes for ab initio total-energy calculations using a plane-wave basis
  set},\ }\href {https://doi.org/10.1103/PhysRevB.54.11169} {\bibfield
  {journal} {\bibinfo  {journal} {Phys. Rev. B}\ }\textbf {\bibinfo {volume}
  {54}},\ \bibinfo {pages} {11169} (\bibinfo {year} {1996})}\BibitemShut
  {NoStop}%
\bibitem [{\citenamefont {Kresse}\ and\ \citenamefont
  {Joubert}(1999)}]{Kresse1999}%
  \BibitemOpen
  \bibfield  {author} {\bibinfo {author} {\bibfnamefont {G.}~\bibnamefont
  {Kresse}}\ and\ \bibinfo {author} {\bibfnamefont {D.}~\bibnamefont
  {Joubert}},\ }\bibfield  {title} {\bibinfo {title} {From ultrasoft
  pseudopotentials to the projector augmented-wave method},\ }\href
  {https://doi.org/10.1103/PhysRevB.59.1758} {\bibfield  {journal} {\bibinfo
  {journal} {Phys. Rev. B}\ }\textbf {\bibinfo {volume} {59}},\ \bibinfo
  {pages} {1758} (\bibinfo {year} {1999})}\BibitemShut {NoStop}%
\bibitem [{\citenamefont {Perdew}\ \emph {et~al.}(1996)\citenamefont {Perdew},
  \citenamefont {Burke},\ and\ \citenamefont
  {Ernzerhof}}]{perdew1996generalized}%
  \BibitemOpen
  \bibfield  {author} {\bibinfo {author} {\bibfnamefont {J.~P.}\ \bibnamefont
  {Perdew}}, \bibinfo {author} {\bibfnamefont {K.}~\bibnamefont {Burke}},\ and\
  \bibinfo {author} {\bibfnamefont {M.}~\bibnamefont {Ernzerhof}},\ }\bibfield
  {title} {\bibinfo {title} {Generalized gradient approximation made simple},\
  }\href {https://doi.org/10.1103/PhysRevLett.77.3865} {\bibfield  {journal}
  {\bibinfo  {journal} {Phys. Rev. Lett.}\ }\textbf {\bibinfo {volume} {77}},\
  \bibinfo {pages} {3865} (\bibinfo {year} {1996})}\BibitemShut {NoStop}%
\bibitem [{\citenamefont {Togo}\ and\ \citenamefont {Tanaka}(2015)}]{phonopy}%
  \BibitemOpen
  \bibfield  {author} {\bibinfo {author} {\bibfnamefont {A.}~\bibnamefont
  {Togo}}\ and\ \bibinfo {author} {\bibfnamefont {I.}~\bibnamefont {Tanaka}},\
  }\bibfield  {title} {\bibinfo {title} {First principles phonon calculations
  in materials science},\ }\href
  {https://doi.org/https://doi.org/10.1016/j.scriptamat.2015.07.021} {\bibfield
   {journal} {\bibinfo  {journal} {Scr. Mater.}\ }\textbf {\bibinfo {volume}
  {108}},\ \bibinfo {pages} {1} (\bibinfo {year} {2015})}\BibitemShut {NoStop}%
\bibitem [{\citenamefont {Barnett}\ and\ \citenamefont
  {Landman}(1993)}]{PhysRevB.48.2081}%
  \BibitemOpen
  \bibfield  {author} {\bibinfo {author} {\bibfnamefont {R.~N.}\ \bibnamefont
  {Barnett}}\ and\ \bibinfo {author} {\bibfnamefont {U.}~\bibnamefont
  {Landman}},\ }\bibfield  {title} {\bibinfo {title} {Born-oppenheimer
  molecular-dynamics simulations of finite systems: Structure and dynamics of
  {{(H$_2$O}$)_2$}},\ }\href {https://doi.org/10.1103/PhysRevB.48.2081}
  {\bibfield  {journal} {\bibinfo  {journal} {Phys. Rev. B}\ }\textbf {\bibinfo
  {volume} {48}},\ \bibinfo {pages} {2081} (\bibinfo {year}
  {1993})}\BibitemShut {NoStop}%
\bibitem [{\citenamefont {Mostofi}\ \emph {et~al.}(2008)\citenamefont
  {Mostofi}, \citenamefont {Yates}, \citenamefont {Lee}, \citenamefont {Souza},
  \citenamefont {Vanderbilt},\ and\ \citenamefont
  {Marzari}}]{mostofi2008wannier90}%
  \BibitemOpen
  \bibfield  {author} {\bibinfo {author} {\bibfnamefont {A.~A.}\ \bibnamefont
  {Mostofi}}, \bibinfo {author} {\bibfnamefont {J.~R.}\ \bibnamefont {Yates}},
  \bibinfo {author} {\bibfnamefont {Y.-S.}\ \bibnamefont {Lee}}, \bibinfo
  {author} {\bibfnamefont {I.}~\bibnamefont {Souza}}, \bibinfo {author}
  {\bibfnamefont {D.}~\bibnamefont {Vanderbilt}},\ and\ \bibinfo {author}
  {\bibfnamefont {N.}~\bibnamefont {Marzari}},\ }\bibfield  {title} {\bibinfo
  {title} {Wannier90: A tool for obtaining maximally-localised wannier
  functions},\ }\href
  {https://doi.org/https://doi.org/10.1016/j.cpc.2007.11.016} {\bibfield
  {journal} {\bibinfo  {journal} {Comput. Phys. Commun.}\ }\textbf {\bibinfo
  {volume} {178}},\ \bibinfo {pages} {685} (\bibinfo {year}
  {2008})}\BibitemShut {NoStop}%
\bibitem [{\citenamefont {Wu}\ \emph {et~al.}(2018{\natexlab{b}})\citenamefont
  {Wu}, \citenamefont {Zhang}, \citenamefont {Song}, \citenamefont {Troyer},\
  and\ \citenamefont {Soluyanov}}]{Wtools}%
  \BibitemOpen
  \bibfield  {author} {\bibinfo {author} {\bibfnamefont {Q.}~\bibnamefont
  {Wu}}, \bibinfo {author} {\bibfnamefont {S.}~\bibnamefont {Zhang}}, \bibinfo
  {author} {\bibfnamefont {H.-F.}\ \bibnamefont {Song}}, \bibinfo {author}
  {\bibfnamefont {M.}~\bibnamefont {Troyer}},\ and\ \bibinfo {author}
  {\bibfnamefont {A.~A.}\ \bibnamefont {Soluyanov}},\ }\bibfield  {title}
  {\bibinfo {title} {Wannier{T}ools : An open-source software package for novel
  topological materials},\ }\href
  {https://doi.org/https://doi.org/10.1016/j.cpc.2017.09.033} {\bibfield
  {journal} {\bibinfo  {journal} {Comput. Phys. Commun.}\ }\textbf {\bibinfo
  {volume} {224}},\ \bibinfo {pages} {405 } (\bibinfo {year}
  {2018}{\natexlab{b}})}\BibitemShut {NoStop}%
\bibitem [{\citenamefont {Sancho}\ \emph {et~al.}(1985)\citenamefont {Sancho},
  \citenamefont {Sancho}, \citenamefont {Sancho},\ and\ \citenamefont
  {Rubio}}]{Sancho1985}%
  \BibitemOpen
  \bibfield  {author} {\bibinfo {author} {\bibfnamefont {M.~P.~L.}\
  \bibnamefont {Sancho}}, \bibinfo {author} {\bibfnamefont {J.~M.~L.}\
  \bibnamefont {Sancho}}, \bibinfo {author} {\bibfnamefont {J.~M.~L.}\
  \bibnamefont {Sancho}},\ and\ \bibinfo {author} {\bibfnamefont
  {J.}~\bibnamefont {Rubio}},\ }\bibfield  {title} {\bibinfo {title} {Highly
  convergent schemes for the calculation of bulk and surface green functions},\
  }\href {https://doi.org/10.1088/0305-4608/15/4/009} {\bibfield  {journal}
  {\bibinfo  {journal} {J. Phys. F}\ }\textbf {\bibinfo {volume} {15}},\
  \bibinfo {pages} {851–858} (\bibinfo {year} {1985})}\BibitemShut {NoStop}%
\bibitem [{\citenamefont {Islam}\ \emph {et~al.}(2021)\citenamefont {Islam},
  \citenamefont {Ghosh}, \citenamefont {Autieri}, \citenamefont {Chowdhury},
  \citenamefont {Bansil}, \citenamefont {Agarwal},\ and\ \citenamefont
  {Singh}}]{Rajibul_PRB2021}%
  \BibitemOpen
  \bibfield  {author} {\bibinfo {author} {\bibfnamefont {R.}~\bibnamefont
  {Islam}}, \bibinfo {author} {\bibfnamefont {B.}~\bibnamefont {Ghosh}},
  \bibinfo {author} {\bibfnamefont {C.}~\bibnamefont {Autieri}}, \bibinfo
  {author} {\bibfnamefont {S.}~\bibnamefont {Chowdhury}}, \bibinfo {author}
  {\bibfnamefont {A.}~\bibnamefont {Bansil}}, \bibinfo {author} {\bibfnamefont
  {A.}~\bibnamefont {Agarwal}},\ and\ \bibinfo {author} {\bibfnamefont
  {B.}~\bibnamefont {Singh}},\ }\bibfield  {title} {\bibinfo {title} {Tunable
  spin polarization and electronic structure of bottom-up synthesized
  {MoSi$_2$N$_4$} materials},\ }\href
  {https://doi.org/10.1103/PhysRevB.104.L201112} {\bibfield  {journal}
  {\bibinfo  {journal} {Phys. Rev. B}\ }\textbf {\bibinfo {volume} {104}},\
  \bibinfo {pages} {L201112} (\bibinfo {year} {2021})}\BibitemShut {NoStop}%
\bibitem [{\citenamefont {van~der Zande}\ \emph {et~al.}(2013)\citenamefont
  {van~der Zande}, \citenamefont {Huang}, \citenamefont {Chenet}, \citenamefont
  {Berkelbach}, \citenamefont {You}, \citenamefont {Lee}, \citenamefont
  {Heinz}, \citenamefont {Reichman}, \citenamefont {Muller},\ and\
  \citenamefont {Hone}}]{SoTMD_Exp2013}%
  \BibitemOpen
  \bibfield  {author} {\bibinfo {author} {\bibfnamefont {A.~M.}\ \bibnamefont
  {van~der Zande}}, \bibinfo {author} {\bibfnamefont {P.~Y.}\ \bibnamefont
  {Huang}}, \bibinfo {author} {\bibfnamefont {D.~A.}\ \bibnamefont {Chenet}},
  \bibinfo {author} {\bibfnamefont {T.~C.}\ \bibnamefont {Berkelbach}},
  \bibinfo {author} {\bibfnamefont {Y.}~\bibnamefont {You}}, \bibinfo {author}
  {\bibfnamefont {G.-H.}\ \bibnamefont {Lee}}, \bibinfo {author} {\bibfnamefont
  {T.~F.}\ \bibnamefont {Heinz}}, \bibinfo {author} {\bibfnamefont {D.~R.}\
  \bibnamefont {Reichman}}, \bibinfo {author} {\bibfnamefont {D.~A.}\
  \bibnamefont {Muller}},\ and\ \bibinfo {author} {\bibfnamefont {J.~C.}\
  \bibnamefont {Hone}},\ }\bibfield  {title} {\bibinfo {title} {Grains and
  grain boundaries in highly crystalline monolayer molybdenum disulphide},\
  }\href {https://doi.org/10.1038/nmat3633} {\bibfield  {journal} {\bibinfo
  {journal} {Nat. Mater.}\ }\textbf {\bibinfo {volume} {12}},\ \bibinfo {pages}
  {554–561} (\bibinfo {year} {2013})}\BibitemShut {NoStop}%
\bibitem [{\citenamefont {Sun}\ \emph {et~al.}(2015)\citenamefont {Sun},
  \citenamefont {Felser},\ and\ \citenamefont {Yan}}]{SOTMD_PRB2015}%
  \BibitemOpen
  \bibfield  {author} {\bibinfo {author} {\bibfnamefont {Y.}~\bibnamefont
  {Sun}}, \bibinfo {author} {\bibfnamefont {C.}~\bibnamefont {Felser}},\ and\
  \bibinfo {author} {\bibfnamefont {B.}~\bibnamefont {Yan}},\ }\bibfield
  {title} {\bibinfo {title} {Graphene-like dirac states and quantum spin hall
  insulators in square-octagonal {M{X}$_{2}$} ({$M=$ Mo, W}; {$X=$ S, Se, Te})
  isomers},\ }\href {https://doi.org/10.1103/PhysRevB.92.165421} {\bibfield
  {journal} {\bibinfo  {journal} {Phys. Rev. B}\ }\textbf {\bibinfo {volume}
  {92}},\ \bibinfo {pages} {165421} (\bibinfo {year} {2015})}\BibitemShut
  {NoStop}%
\bibitem [{\citenamefont {Fu}\ and\ \citenamefont {Kane}(2007)}]{Fu2007}%
  \BibitemOpen
  \bibfield  {author} {\bibinfo {author} {\bibfnamefont {L.}~\bibnamefont
  {Fu}}\ and\ \bibinfo {author} {\bibfnamefont {C.~L.}\ \bibnamefont {Kane}},\
  }\bibfield  {title} {\bibinfo {title} {Topological insulators with inversion
  symmetry},\ }\href {https://doi.org/10.1103/PhysRevB.76.045302} {\bibfield
  {journal} {\bibinfo  {journal} {Phys. Rev. B}\ }\textbf {\bibinfo {volume}
  {76}},\ \bibinfo {pages} {045302} (\bibinfo {year} {2007})}\BibitemShut
  {NoStop}%
\bibitem [{\citenamefont {Wang}\ \emph {et~al.}(2023)\citenamefont {Wang},
  \citenamefont {Hung}, \citenamefont {Zhou}, \citenamefont {Ong},\ and\
  \citenamefont {Lin}}]{FST_Wang2023}%
  \BibitemOpen
  \bibfield  {author} {\bibinfo {author} {\bibfnamefont {B.}~\bibnamefont
  {Wang}}, \bibinfo {author} {\bibfnamefont {Y.-C.}\ \bibnamefont {Hung}},
  \bibinfo {author} {\bibfnamefont {X.}~\bibnamefont {Zhou}}, \bibinfo {author}
  {\bibfnamefont {T.}~\bibnamefont {Ong}},\ and\ \bibinfo {author}
  {\bibfnamefont {H.}~\bibnamefont {Lin}},\ }\bibfield  {title} {\bibinfo
  {title} {Feature spectrum topology},\ }\href
  {https://arxiv.org/abs/2310.14832} {\  (\bibinfo {year} {2023})},\ \Eprint
  {https://arxiv.org/abs/2310.14832} {arXiv:2310.14832} \BibitemShut {NoStop}%
\bibitem [{\citenamefont {Yao}\ \emph {et~al.}(2024)\citenamefont {Yao},
  \citenamefont {Zhou}, \citenamefont {Hung}, \citenamefont {Lin},
  \citenamefont {Bansil},\ and\ \citenamefont {Chang}}]{PhysRevB.109.155143}%
  \BibitemOpen
  \bibfield  {author} {\bibinfo {author} {\bibfnamefont {Y.-T.}\ \bibnamefont
  {Yao}}, \bibinfo {author} {\bibfnamefont {X.}~\bibnamefont {Zhou}}, \bibinfo
  {author} {\bibfnamefont {Y.-C.}\ \bibnamefont {Hung}}, \bibinfo {author}
  {\bibfnamefont {H.}~\bibnamefont {Lin}}, \bibinfo {author} {\bibfnamefont
  {A.}~\bibnamefont {Bansil}},\ and\ \bibinfo {author} {\bibfnamefont {T.-R.}\
  \bibnamefont {Chang}},\ }\bibfield  {title} {\bibinfo {title} {Feature-energy
  duality of topological boundary states in a multilayer quantum spin hall
  insulator},\ }\href {https://doi.org/10.1103/PhysRevB.109.155143} {\bibfield
  {journal} {\bibinfo  {journal} {Phys. Rev. B}\ }\textbf {\bibinfo {volume}
  {109}},\ \bibinfo {pages} {155143} (\bibinfo {year} {2024})}\BibitemShut
  {NoStop}%
\bibitem [{\citenamefont {Sun}\ \emph {et~al.}(2016)\citenamefont {Sun},
  \citenamefont {Zhang}, \citenamefont {Felser},\ and\ \citenamefont
  {Yan}}]{SHC_TaAs}%
  \BibitemOpen
  \bibfield  {author} {\bibinfo {author} {\bibfnamefont {Y.}~\bibnamefont
  {Sun}}, \bibinfo {author} {\bibfnamefont {Y.}~\bibnamefont {Zhang}}, \bibinfo
  {author} {\bibfnamefont {C.}~\bibnamefont {Felser}},\ and\ \bibinfo {author}
  {\bibfnamefont {B.}~\bibnamefont {Yan}},\ }\bibfield  {title} {\bibinfo
  {title} {Strong intrinsic spin hall effect in the {TaAs} family of weyl
  semimetals},\ }\href {https://doi.org/10.1103/PhysRevLett.117.146403}
  {\bibfield  {journal} {\bibinfo  {journal} {Phys. Rev. Lett.}\ }\textbf
  {\bibinfo {volume} {117}},\ \bibinfo {pages} {146403} (\bibinfo {year}
  {2016})}\BibitemShut {NoStop}%
\bibitem [{\citenamefont {Qiao}\ \emph {et~al.}(2018)\citenamefont {Qiao},
  \citenamefont {Zhou}, \citenamefont {Yuan},\ and\ \citenamefont
  {Zhao}}]{SHC_w90}%
  \BibitemOpen
  \bibfield  {author} {\bibinfo {author} {\bibfnamefont {J.}~\bibnamefont
  {Qiao}}, \bibinfo {author} {\bibfnamefont {J.}~\bibnamefont {Zhou}}, \bibinfo
  {author} {\bibfnamefont {Z.}~\bibnamefont {Yuan}},\ and\ \bibinfo {author}
  {\bibfnamefont {W.}~\bibnamefont {Zhao}},\ }\bibfield  {title} {\bibinfo
  {title} {Calculation of intrinsic spin hall conductivity by wannier
  interpolation},\ }\href {https://doi.org/10.1103/PhysRevB.98.214402}
  {\bibfield  {journal} {\bibinfo  {journal} {Phys. Rev. B}\ }\textbf {\bibinfo
  {volume} {98}},\ \bibinfo {pages} {214402} (\bibinfo {year}
  {2018})}\BibitemShut {NoStop}%
\bibitem [{\citenamefont {Yuan}\ \emph {et~al.}(2019)\citenamefont {Yuan},
  \citenamefont {Isobe},\ and\ \citenamefont {Fu}}]{Yuan2019}%
  \BibitemOpen
  \bibfield  {author} {\bibinfo {author} {\bibfnamefont {N.~F.~Q.}\
  \bibnamefont {Yuan}}, \bibinfo {author} {\bibfnamefont {H.}~\bibnamefont
  {Isobe}},\ and\ \bibinfo {author} {\bibfnamefont {L.}~\bibnamefont {Fu}},\
  }\bibfield  {title} {\bibinfo {title} {Magic of high-order van hove
  singularity},\ }\href {http://dx.doi.org/10.1038/s41467-019-13670-9}
  {\bibfield  {journal} {\bibinfo  {journal} {Nat. Commun.}\ }\textbf {\bibinfo
  {volume} {10}} (\bibinfo {year} {2019})}\BibitemShut {NoStop}%
\end{thebibliography}%

\renewcommand{\thefigure}{S\arabic{figure}}
\renewcommand{\thesection}{S-\Roman{section}}
\renewcommand{\thetable}{S\Roman{table}}

\setcounter{figure}{0}
\setcounter{section}{0}
\clearpage
\title{--Supplementary Materials-- \\Emergent spin Hall quantization and high-order van Hove singularities in square-octagonal MA$_2$Z$_4$}
\maketitle
\onecolumngrid

\begin{figure}[htp]
\includegraphics[width=0.70 \linewidth]{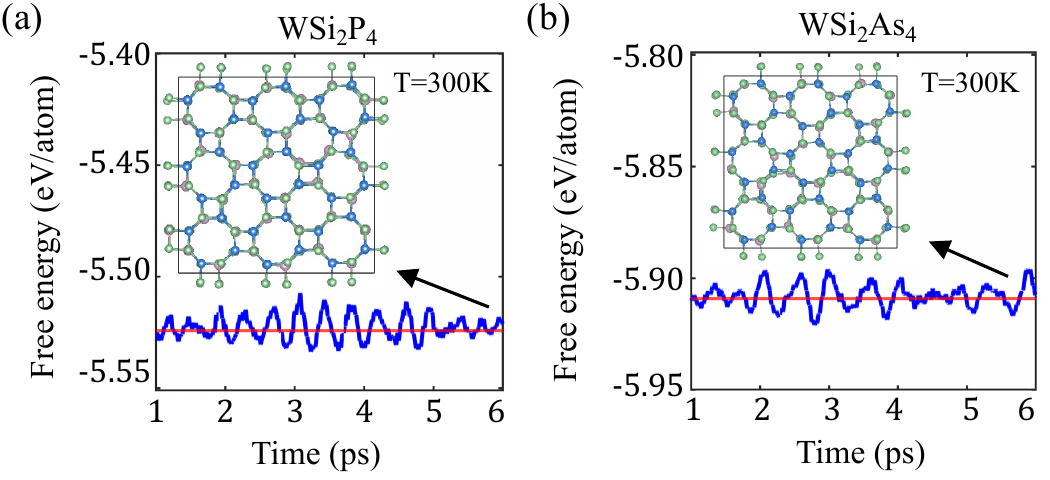}
\caption{\textbf{Thermodynamical stability of square-octagonal (SO) monolayers.} Free energy as a function of simulation time during ab initio molecular dynamics simulations at $T=300$~K for monolayers (a) WSi$_2$P$_4$ and (b) WSi$_2$As$_4$. The red line represents the mean free energy per atom. Both monolayers exhibit only small fluctuations around their respective ground-state energies throughout the simulation. The insets show the relaxed supercell structures at the end of the simulation, confirming that no bond breaking or structural distortions occur, demonstrating the thermodynamical stability at room temperature.}
\label{structure}
\end{figure}

\begin{figure}[htp]
\includegraphics[width=1.0\linewidth]{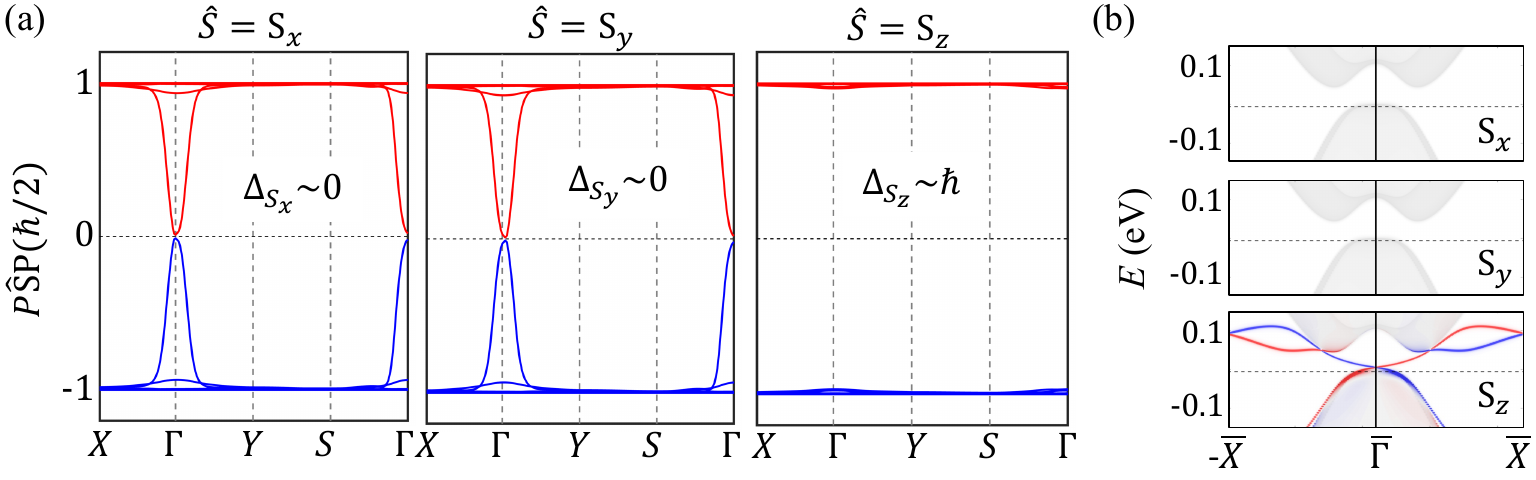}
\caption{\textbf{Spin-feature spectrum and edge states.}(a) Spin-resolved feature spectrum ($P\hat{S}P$) of the bulk occupied bands of WSi$_2$As$_4$ for spin components $\hat{S}=S_x$, $S_y$, and $S_z$. The spectrum is gapless for $S_x$ and $S_y$ near band inversion $\Gamma$ point, while it is gapped for $S_z$ with eigenvalues pinned near $\pm \hbar/2$ with slight deviations. The spin gap $\Delta_{S_x}\sim 0$, $\Delta_{S_y}\sim 0$ and $\Delta_{S_z}\sim \hbar$, indicates approximate spin-rotational symmetry about the $z$-axis consistent with a spin U(1) quasi-symmetry. (b) The (010) spin-resolved edge band structure of WSi$_2$As$_4$, shows $S_z$-polarized edge states with negligible contribution from spin $S_x$ and $S_y$ components.}
\label{spinspectrum}
\end{figure}

\begin{figure}[htp]
\includegraphics[width=1.0\linewidth]{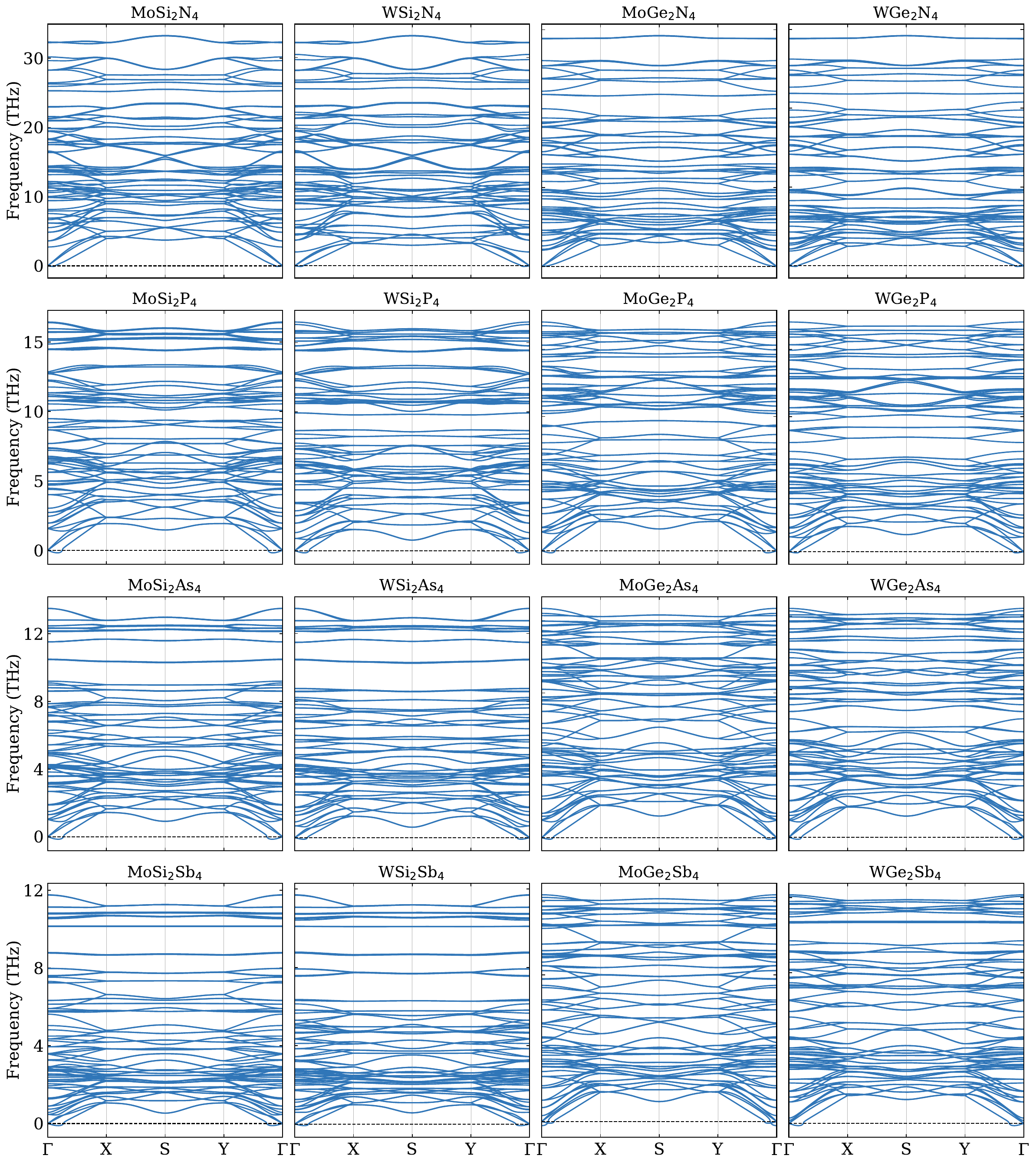}
\caption{\textbf{Phonon spectra.} Phonon spectrum of monolayer MA$_2$Z$_4$ ($M=$ Mo/W ; $A =$ Si/Ge ; $Z=$ Pnictogen) compounds. The spectra exhibit no imaginary modes throughout the Brillouin zone, indicating the dynamical stability of all SO monolayer compounds.}
\label{phonons}
\end{figure}

\begin{figure}[htp]
\includegraphics[width=0.95\linewidth]{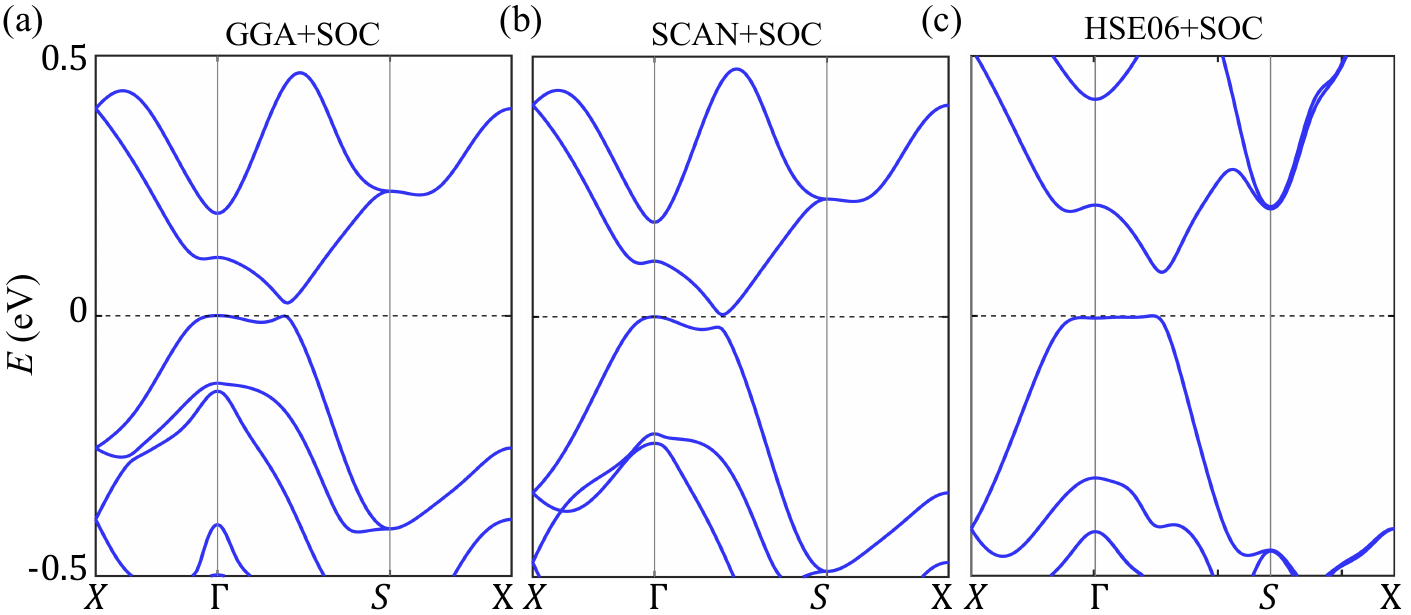}
\caption{\textbf{Topology and band features of WSi$_2$Sb$_4$ with different functionals.} Band structure of monolayer SO WSi$_2$Sb$_4$ calculated using (a) GGA, (b) SCAN, and (c) HSE06 hybrid exchange--correlation functionals, including spin--orbit coupling. The band topology, quasi-flat band along the $\Gamma$--$S$ direction, and associated van Hove singularities (VHSs) near the Fermi level are robust across all functionals, indicating that these features are intrinsic to the electronic structure. The HSE06 yields a larger band gap and enhances the flatness of the valence band along the $\Gamma$--$S$ direction, indicating stronger localization of electronic states and a greater susceptibility to correlation-driven instabilities.}
\label{phonons}
\end{figure}

\renewcommand{\arraystretch}{1.5}
\begin{table*}[ht!]
\caption{Total relative energies (in meV/atom) of various polymorphs of MA$_2$Z$_4$ compounds ($M$ = Mo, W; $A$ = Si, Ge; $Z$ = Pnictogen), calculated with respect to the 1H phase. The energies are normalized per atom, and the lowest-energy (ground state) phase for each composition is highlighted in bold.}
\centering
\begin{tabularx}{0.95\textwidth}{Y Y Y Y  Y}
\hline\hline
Material & $\Delta$E$_{1H}$ & $\Delta$E$_{1T}$ & $\Delta$E$_{1T^\prime}$ & $\Delta$E$_{SO}$ \\
\hline
MoSi$_2$N$_4$       & \textbf{0} & 77 		& 70 		& 151 \\
WSi$_2$N$_4$         & \textbf{0} & 85 		& 75 		& 164 \\
MoGe$_2$N$_4$      & \textbf{0} & 67 		& 54 		& 133 \\
WGe$_2$N$_4$       & \textbf{0} & 73 		& 55 		& 144 \\
MoSi$_2$P$_4$        & \textbf{0} & 46		 & 29 		& 85 \\
WSi$_2$P$_4$         & \textbf{0} & 47 		& 23 		& 87 \\
MoGe$_2$P$_4$      & \textbf{0} & 46 		& 22 		& 79 \\
WGe$_2$P$_4$        & \textbf{0} & 45 		& 16 		& 82 \\
MoSi$_2$As$_4$      & \textbf{0} & 36 		& 14 		& 72 \\
WSi$_2$As$_4$        & \textbf{0} & 37 		& 5 		& 72 \\
MoGe$_2$As$_4$     & \textbf{0} & 34 		& 7		 & 70 \\
WGe$_2$As$_4$      & 0 & 33 		&\textbf{-1} 		& 69 \\
MoSi$_2$Sb$_4$      & 0 & 22		 & \textbf{-2} 		& 59 \\
WSi$_2$Sb$_4$       & 0 & 22 		&\textbf{-13} 		& 55 \\
MoGe$_2$Sb$_4$    & 0 & 21 		& \textbf{-7} 		& 57 \\
WGe$_2$Sb$_4$      & 0 & 19 		& \textbf{-15}		 & 54 \\
\hline\hline
\end{tabularx}
\label{table1}
\end{table*}

\begin{figure}[htp]
\includegraphics[width=1.0\linewidth]{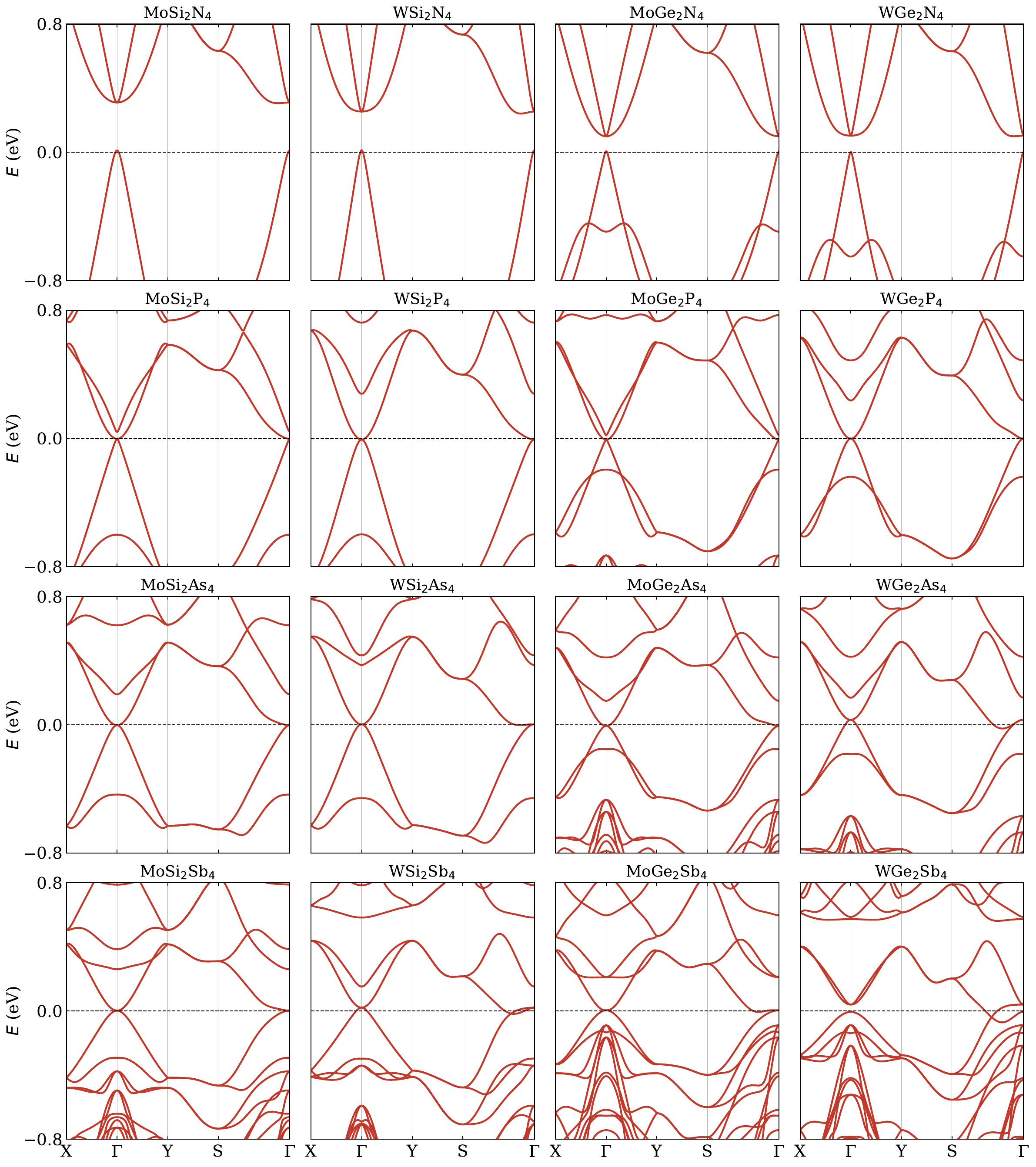}
\caption{\textbf{Band structure without spin--orbit coupling.} Bulk band structure of SO monolayer MA$_2$Z$_4$ (M = Mo/W ; A = Si/Ge ; Z = pnictogen) compounds in absence of spin-orbit coupling using GGA. Materials with $Z =$ N is insulator, while $Z =$ P, As, Sb shows semi-metallic character with parabolic-like dispersion near the Ferm level at $\Gamma$ point. Notably, MoGe$_2$Sb$_4$ and WGe$_2$Sb$_4$ show metallic behavior, highlighting the diversity of electronic phases in these compounds.}
\label{bands_wso}
\end{figure}

\begin{figure}[htp]
\includegraphics[width=1.0\linewidth]{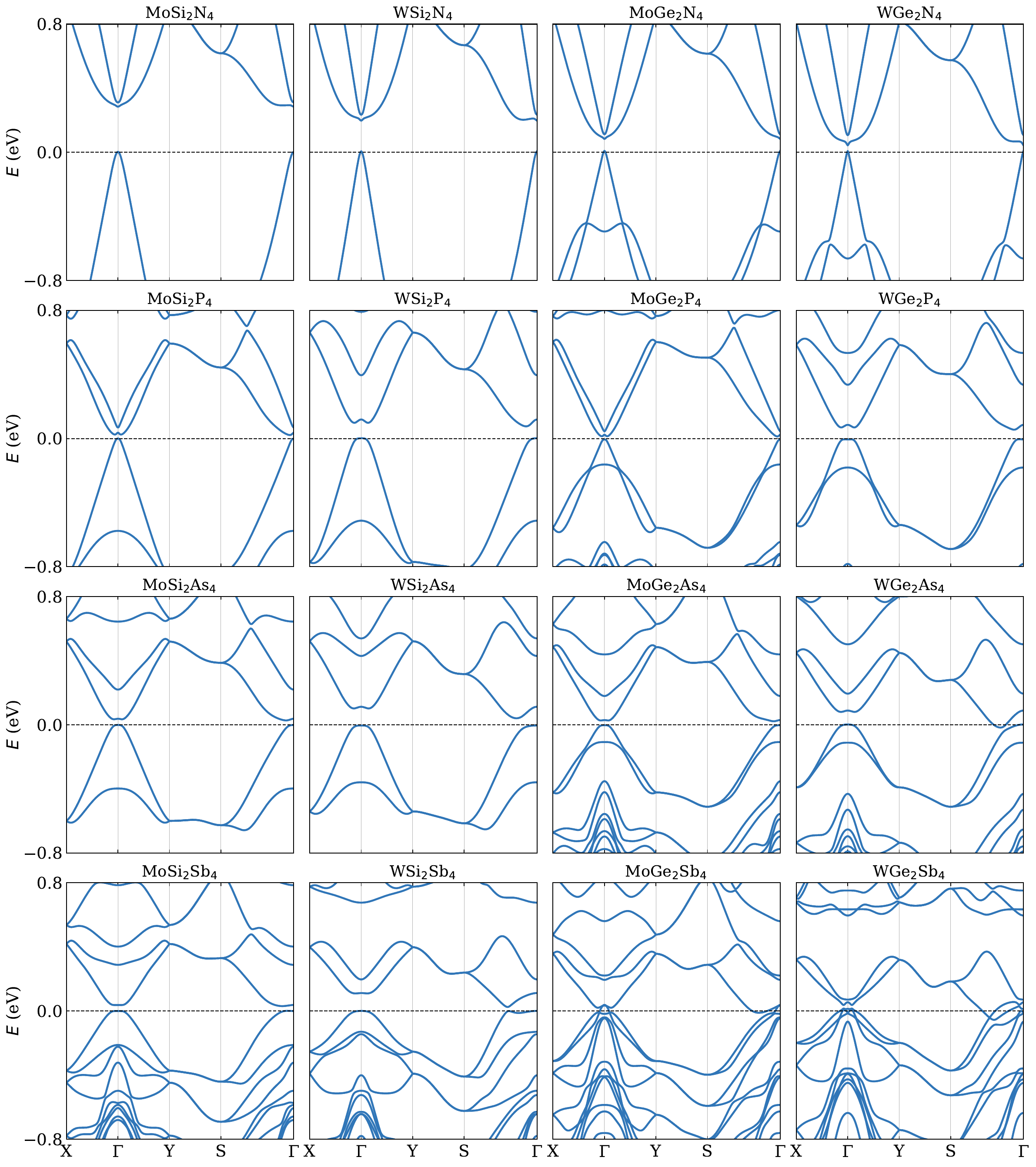}
\caption{\textbf{Band structure with spin--orbit coupling.} Bulk band structure of SO monolayer MA$_2$Z$_4$ (M = Mo/W ; A = Si/Ge ; Z = pnictogen) compounds in presence of spin-orbit coupling using GGA. 
The inclusion of SOC lifts spinless degeneracies and realize a quantum spin Hall insulator state with $\mathbb{Z}_2=1$ and spin-Chern number $C_s=1$ for $Z =$P, As, and Sb, while for $Z=$N remains $\mathbb{Z}_2=0$ trivial insulator.}
\label{bands_so}
\end{figure}

\end{document}